\documentclass[journal,draftcls,onecolumn,12pt]{IEEEtran}
\usepackage{tablefootnote}
\usepackage{makecell}
\usepackage{tabularx}
\usepackage{array}
\usepackage{graphics} 
\graphicspath{{./picture/}}
\usepackage{float}
\usepackage{tikz}
\usepackage{xcolor} 
\usepackage{lipsum} 
\usepackage{epsfig} 
\usepackage{cite}
\usepackage{amsmath} 
\usepackage{amssymb}  
\usepackage{bm}
\usepackage{changepage} 
\usepackage{algorithm}  
\usepackage{latexsym}
\usepackage{algorithmicx}
\usepackage{algpseudocode}
\usepackage{graphicx} 
\usepackage{epstopdf}
\usepackage[short]{optidef}
\usepackage{hyperref}
\hypersetup{linkbordercolor=pink}

\newtheorem{remark}{Remark}

\begin{document}

\title{ Deep Reinforcement Learning for Multi-user Massive MIMO with Channel Aging
}

\author{

\IEEEauthorblockN{Zhenyuan Feng,  \IEEEmembership{ Member, IEEE}, Bruno Clerckx, \IEEEmembership{Fellow, IEEE}}
 \\\vspace{-4mm}
\thanks{Z. Feng is with the Department of Electrical and Electronic Engineering,
Imperial College London, London SW7 2AZ, U.K. (e-mail: z.feng19@imperial.ac.uk).B. Clerckx is with the Department of Electrical and Electronic Engineering at Imperial College London, London SW7 2AZ, UK and with Silicon Austria Labs (SAL), Graz A-8010, Austria (email: b.clerckx@imperial.ac.uk; bruno.clerckx@silicon-austria.com)}
}

\maketitle
\begin{abstract}
The design of beamforming for downlink  multi-user massive multi-input multi-output (MIMO) relies on accurate downlink channel state information (CSI) at the transmitter (CSIT). In fact, it is difficult for the base station (BS) to obtain perfect CSIT due to user mobility, {and} latency/feedback delay (between downlink data transmission and CSI acquisition). Hence, robust beamforming under imperfect CSIT is needed. In this paper, considering multiple antennas at all nodes (base station and user terminals), we develop a multi-agent deep reinforcement learning (DRL) framework for massive MIMO under imperfect CSIT, where the transmit and receive beamforming are jointly designed  to maximize the average information rate of all users. Leveraging this DRL-based framework, interference management is explored and  three DRL-based  schemes, namely the distributed-learning-distributed-processing scheme, partial-distributed-learning-distributed-processing, and  central-learning-distributed-processing scheme, are proposed and analyzed. This paper \textrm{1)} highlights the fact that the DRL-based strategies  outperform the random action-chosen strategy and the delay-sensitive strategy named as sample-and-hold (SAH) approach, 
and  achieved over 90$\%$ of the information rate of two selected benchmarks with lower complexity: the zero-forcing channel-inversion (ZF-CI)  with perfect CSIT and the Greedy Beam Selection strategy, 
\textrm{2)} demonstrates the inherent robustness of the proposed designs in the presence of {channel aging.  \textrm{3)} conducts detailed convergence and scalability analysis on the proposed framework.}
\end{abstract}


\section{Introduction}
\IEEEPARstart{D}{ue} to the increasing demand for data and connectivity in fifth-generation (5G) \cite{andrews2014will} and sixth-generation (6G) \cite{dang2020should}, multi-antenna technologies have attracted great attention in academia and industry. The research on multi-antenna techniques has  promoted   the development of {multi-input multi-output
(MIMO)}  technology. MIMO nowadays plays an indispensable role in the physical layer, media access control (MAC) layer, and network layer in  wireless communications and networking \cite{xiao2021antenna}.  {At the physical layer, multi-antenna beamforming strategies have attracted great interest  due to their ability to achieve considerable antenna gains, multiplexing gains, and diversity gains \cite{stankovic2008generalized, clerckx2013mimo}, and gradually  evolved into a massive MIMO system, in which the number of antennas at the BS reaches tens or even hundreds, attracting a larger number of users.  To enable a high throughput in the   massive MIMO  system, the base station (BS) relies on the huge demand for global and instantaneous channel state information (CSI) based on efficient channel estimation techniques \cite{ding2014compressive, meng2011compressive}.}
Nevertheless, the ground/air/space platforms such as high-speed trains/unmanned aerial vehicles (UAV)/satellites have a common characteristic of 3D mobility which leads to  a  stringent time constraint on CSI acquisition and even causes misalignment of narrow beams. 
Therefore, in  future communication systems, how to maintain good connectivity and system capacity without perfect {channel state information at the transmitter (CSIT)} (so-called imperfect CSIT) is regarded as an important problem that yearns for prompt solutions.

The imperfect CSIT is usually caused by the drastic change of the propagation environment due to user mobility \cite{truong2013effects} and CSI feedback/acquisition delay between the base station (BS) and users \cite{ramya2009using}. {The CSI delay due to user mobility or feedback or acquisition delay} is the time gap between the time point when {the downlink training happens} and the BS starts downlink data transmission with the estimated
channel. Such delay can be in the level of milliseconds which causes the estimated channels to be outdated when actually downlink transmission happens. This delay becomes more catastrophic at high user mobility since rapid channel variation inevitably  causes  performance degradation in massive MIMO systems\cite{papazafeiropoulos2016impact}. {This phenomenon, which is known as \textit{channel aging}, describes the mismatch between estimated and update-to-date channels. That is to say, it represents the divergence arising between the channel estimation happening in the BS/users and the actual channel through which the data transmission occurs.}

\subsection{{Related Works}}
 {To address the issue above, many papers have studied massive MIMO systems with imperfect CSIT \cite{truong2013effects, ramya2009using, papazafeiropoulos2016impact, lee2013space, lee2014distributed, yin2020addressing, papazafeiropoulos2014linear,kong2015sum,9491092} since CSIT is pivotal to the performance of systems that account for a great number of antennas and users. In \cite{truong2013effects}, the impact of channel aging due to mobility is partially overcome through finite impulse response Wiener predictor without considering hardware phase noise, which is further studied in \cite{papazafeiropoulos2016impact}. To tackle the CSI feedback/acquisition delay,}
 one strategy is to use space-time interference alignment to optimize the degree of freedom (DoF) with delayed CSIT \cite{lee2013space, lee2014distributed}. Another method investigates the channel prediction based on  {the} channel correlation \cite{yin2020addressing} and past CSI \cite{papazafeiropoulos2014linear}.  {In addition, to maintain the multi-user connectivity and mitigate the degrading effect of user mobility, low complexity power allocation methods are derived in \cite{kong2015sum} for Space Division Multiple Access (SDMA) which is outperformed by Rate-Splitting Multiple Access (RSMA) in \cite{9491092} in terms of ergodic sum-rate.}

 {On the one hand, the channel prediction approaches in the papers cited above \cite{truong2013effects, ramya2009using, papazafeiropoulos2016impact, lee2013space, lee2014distributed, yin2020addressing, papazafeiropoulos2014linear} demonstrate good performance but experience extremely high complexity in channel prediction algorithms due to  the increasing dimension of the antenna arrays. On the other hand, power allocation strategies in \cite{kong2015sum,9491092} exhibit lower complexity  but sacrifice performance for tractability. To maintain a better balance of performance and complexity}, an alternative  strategy with lower complexity and looser CSI requirement needs to be developed urgently.

Machine learning (ML) \cite{anzai2012pattern} has demonstrated  great usefulness in wireless  systems \cite{ nerini2022machine, ma2022reinforcement,kim2021learning, lin2019beamforming, yuan2020machine, kim2020massive, wu2021channel, qin2022partial, zhang2022predicting,nasir2019multi, ge2020deep, zhang2020deep, huang2020reconfigurable,  yang2020deep,  huang2021multi, li2021deep, zhang2021joint, chen2021beamforming,hu2021joint, ren2022long, fozi2021fast}. To cope with complex problems in a large-dimensional
 MIMO system, deep learning (DL) {has drawn} research {interest} in not only  beamforming design \cite{kim2021learning, lin2019beamforming} by feeding  CSI to the neural network but also channel prediction \cite{yuan2020machine, kim2020massive, wu2021channel, qin2022partial, zhang2022predicting} by treating the time-varying channel as a time series,
 thanks to the strong representation capability of the deep neural network (DNN). Nevertheless, under  stringent time constraints  in mobility scenarios, 
the excellent generalization performance of DNN can not be fully exploited due to an insufficient number of data samples. 
In view of it, by elaborately  treating the time-varying channel problem as a Markov decision process (MDP), deep reinforcement learning (DRL) has been regarded as a useful technology  {to design wireless communication systems by leveraging  fast convergence of DL frameworks as well as continuous improvement characteristic in reinforcement learning (RL) algorithms  \cite{luong2019applications, nasir2019multi, ge2020deep, zhang2020deep, huang2020reconfigurable,  yang2020deep,  huang2021multi, li2021deep,  chen2021beamforming, zhang2021joint,hu2021joint, ren2022long, fozi2021fast}}. Systematically, a comprehensive  tutorial in \cite{luong2019applications} reveals the applications of DRL for 5G and beyond.  {DRL is used to solve the power allocation problem in time-varying channels in  \cite{nasir2019multi} for single transmit antenna scenarios, and is further studied in \cite{ge2020deep} for multi-antenna beamforming  and  in \cite{chen2021beamforming} for multi-user conditions. In \cite{huang2020reconfigurable, yang2020deep, huang2021multi, ren2022long}, DRL is utilized to tackle the passive beamforming design problem in reconfigurable intelligent surfaces (RIS)-aided communications and help reduce the computations compared to alternative frameworks.}
In terms of active beamforming using DRL, several efforts have been made on designing low complexity algorithms based on  deep Q-network (DQN) \cite{ge2020deep, chen2021beamforming, zhang2021joint, hu2021joint} and partially observed MDP \cite{fozi2021fast} frameworks. 
\subsection{{Motivation and Specific Contributions of the Paper}}
\begin{figure}[t]
    \centering
    \includegraphics[width = 8cm]{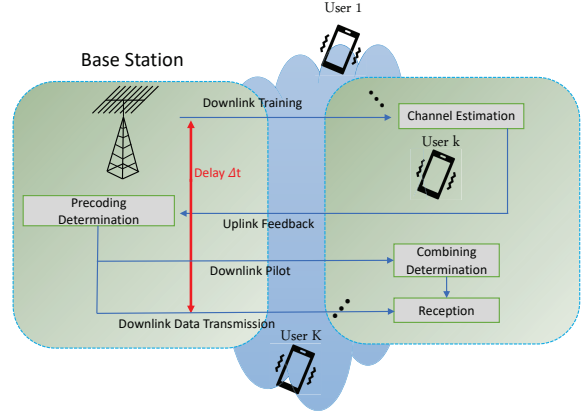}%
    \caption{ The system model for FDD-based pilot process.  The CSI feedback or acquisition delay $\Delta t$ is the time gap between the time point when the
channel is estimated and the BS starts downlink data transmission with the estimated channel.}
   \label{fig:ZFCI}
\end{figure}

Existing works \cite{nasir2019multi, ge2020deep, zhang2020deep, huang2020reconfigurable,  yang2020deep, huang2021multi, li2021deep, chen2021beamforming,zhang2021joint, hu2021joint,ren2022long} assume that perfect CSIT or instantaneous channel gain via receiver feedback is  known  at the transmitter. Unfortunately, such an assumption is   impractical in real-world systems with CSI feedback/acquisition delay and user mobility \cite{truong2013effects,ramya2009using}.  In addition, beamforming is not limited to the transmitter and can also be used at the receiver to perform better interference management. To our best knowledge, predicting the beamformers of both transmitter and receiver with imperfect CSIT is never considered in DRL-based papers. Instead, all the existing work focuses  on high-level multi-cell single-user (SU) single-input-single-output (SISO) \cite{nasir2019multi, zhang2020deep} (no transmit and receive beamforming) and multi-input-single-output (MISO) (only transmit beamforming) \cite{ge2020deep, yang2020deep, zhang2021joint, fozi2021fast} scenarios without considering  multiple receive antenna cases, which motivates this work. 
{In addition to DRL-based strategies, a traditional procedure of beamforming is to use the frequency-division duplexing (FDD)  pilot-based channel estimation procedure and  zero-forcing channel inversion (ZF-CI) scheme which is shown in Fig. \ref{fig:ZFCI} \cite{stankovic2008generalized}. Compared with DRL-based approaches, a key disadvantage of this method is that the system performance is heavily dependent on how fast the channel is changing as well as the feedback delay.} 

{Motivated by the above}, we study  the joint transmit precoder and receive combiner design  in massive MIMO downlink transmission with {channel aging}. The contributions of this paper are summarized as follows.
\begin{itemize}
    \item We construct an efficient multi-agent DRL-based framework for massive MIMO downlink transmission\footnote{The terminology massive MIMO in this paper implies multiple receive antennas}, { in light of which three DRL-based algorithms were derived based on stream-level, user-level, and system-level agent modeling. This is the first paper showing that the DRL-based framework can be used to address very high-dimension optimization problems and demonstrates 1) robustness on the degrading effect of channel aging, 2) stringent interference management especially the inter-stream interference and multi-user interference.  }
    \item To address the challenge of high-dimensional antenna beamforming problems, by utilizing the DRL-based framework, three  DRL-based schemes, namely distributed-learning-distributed-processing DRL-based  scheme (DDRL), partial-distributed-learning-distributed-processing DRL-based scheme (PDRL), and central-learning-distributed-processing DRL-based scheme (CDRL), are proposed, analyzed,  and evaluated. For  DDRL, each stream is modeled as an agent. All the agents save their experiences {in} a private experience pool for later training. In contrast, in CDRL, the whole system is modeled as a central agent. What's more, to bridge  DDRL and CDRL, we demonstrate another algorithm, i.e., PDRL which offers a more flexible design by modeling each user as an agent  to balance the performance and complexity. Note that the DDRL and CDRL are different from those in \cite{nasir2019multi, ge2020deep} since we are tackling the problem with 1) receive beamforming with multiple receive antennas, 2) transmit beamforming under imperfect CSIT, 3) multiple streams for each user, and 4) {a large number of transmit antennas at BS} compared with SISO in \cite{nasir2019multi} and {4-antenna} MISO in \cite{ge2020deep, chen2021beamforming}, respectively. 
    
    \item Leveraging the DRL-based framework  mentioned above,  the precoders at BS and combiners at users are jointly designed  by gradually maximizing the average information rate through the observed  reward. In particular, the BS decides the transmit precoder and receive combiner for each stream with imperfect CSIT and perfect CSIR. The merits of this design are shown through extensive simulations  by benchmarking our schemes against the conventional, sample-and-hold (SAH) approach  \cite{zhang2022predicting}, zero-forcing channel-inversion (ZF-CI) strategy \cite{stankovic2008generalized},   {greedy beam selection} and random action-chosen scheme.
   
    \item We demonstrate the advantages of DRL-based strategies  over the  benchmarks above. In particular, the  proposed algorithms {show} \textrm{1)}   fast convergence to efficient beamforming policy, \textrm{2)} the robustness on tracing the channel dynamic against channel uncertainty {due to channel aging}, and \textrm{3)} lower complexity compared with traditional beamforming strategy. All of these properties are essential in practical wireless networks.
    \item By numerical results,  we show that  our proposed DRL-based schemes outperform the SAH approach  and  random action-chosen scheme. In particular, DDRL can achieve nearly 90$\%$ of  the performance of the state-of-the-art ZF-CI method with perfect CSI (ZF-CI PCSI) and $95\%$ of the performance of the Greedy Beam Selection method {but incurs more hardware complexity and more uplink overhead in an FDD setup}. By increasing the resolution of the codebook  and hyper-parameter tuning on the reward function, the performance can be further improved.
\end{itemize}

\textit{Organizations:} The whole Section \ref{section2} is devoted to the system model, channel  model, and  the formulated sum-rate problem. In Section \ref{section4}, the basics of DRL are introduced, and three practical multi-agent DRL-based  approaches are proposed. The simulation results are demonstrated in  Section \ref{Section5} and this paper is concluded in Section \ref{section6}.

\textit{Notations:} Boldface lower- and upper-case letters $\mathbf{H}$, and $\mathbf{h}$, denote vectors and matrices, respectively. $\mathbb{E}\{\cdot\}$ represents  statistical expectation. $(\cdot)^{-1}, (\cdot)^T, (\cdot)^*$, and $(\cdot)^{H}$ indicate inversion, transpose, conjugate, conjugate-transpose, respectively. $\mathcal{R}$ and $\mathcal{I}$ denote the real and imaginary parts of a complex number, respectively. $\mathbf{I}_M$ denotes an $M\times M$ identity matrix. $\mathbf{0}$ denotes an all-zero matrix. $||\mathbf{a}||$ denotes the norm of a vector $\mathbf{a}$.  $|{a}|$ denotes the norm of a variable $a$. 

\section{System Model}\label{section2}
Consider the MIMO broadcast channel (BC) with  one $M$-antenna BS  and $K$ $N$-antenna users indexed by $\mathcal{K} = \{1, \dots, K\}$\cite{sung2009generalized} . The BS aims to deliver $M_s$ streams in the time instant of interest. For simplicity, {A number of $KN_s$ streams are transmitted simultaneously from the $M$ antennas of the BS. Each group of $N_s$ streams indexed by $N_s \in \mathcal{N} = \{1, \ldots, N_s\}$ is targeted at one of the $K$ users. Note that we consider a setting where $M \geq KN_s$ to ensure the spatial multiplexing gain}. The transmit power $P$ is uniformly allocated to all {$KN_s$}  streams. We assume
that the BS and all users operate in the same time-frequency resource and are synchronized. The transmitted signal, i.e., the precoded data vector,  at time slot $t$ can be written as
\begin{equation}
    \mathbf{x}(t) = \sqrt{\frac{P}{KN_s}}\sum_{k = 1}^{K} \sum_{n = 1}^{N_s}\mathbf{p}_{k,n}(t)s_{k,n}(t)
\end{equation}
where $s_{k,n} , \forall k \in \mathcal{K}, \forall n \in \mathcal{N}, 
$ is the encoded message from message $W_{k, n}$ with zero mean and  $\mathbb{E}(|s_{k,n}|^2) = 1$, and precoder $\mathbf{p}_{k,n}(t)\in\mathbb{C}^{M\times 1}$ is subject to $\Vert\mathbf{p}_{k,n}(t)\Vert^2 = 1$. The received signal at user $k$ can be expressed as
\begin{equation}
    \begin{array}{lr}
    \mathbf{y}_k(t) =\sqrt{\frac{P}{KN_s}} \mathbf{H}_k(t)\sum_{n = 1}^{N_s}\mathbf{p}_{k,n}(t)s_{k,n}(t)\\ + \sqrt{\frac{P}{KN_s}}\mathbf{H}_k(t)\sum_{j \neq k, j = 1}^K\sum_{i = 1}^{N_s}\mathbf{p}_{j,i}(t)s_{j,i}(t) + \mathbf{n}_k(t)
    \end{array}
\end{equation}
where the noise vector $\mathbf{n}_k\in\mathbb{C}^{N\times 1}$ is assumed to follow a complex normal distribution, i.e., $\mathbf{n}_k\in \mathcal{CN}(\mathbf{0}, \sigma_n^2\mathbf{I}_{N})$. At the user side, the combiner vector for each stream is denoted as  $\mathbf{w}_{k,n}(t)\in\mathbb{C}^{N\times 1}$, $\Vert\mathbf{w}_{k,n}(t)\Vert^2 = 1, \forall k \in \mathcal{K}, n \in \mathcal{N}$. Then, the  achievable rate for  user $k$ and {the} average user  rate at time slot $t$ can be written as
\begin{equation}
    R_{k}(t) = \sum_{n = 1}^{N_s} G_{k,n}(t),\quad \bar{R}(t) = \frac{\sum_{k = 1}^{K}R_k(t)}{K}\label{for:totalr}
\end{equation}
, where $G_{k,n}$ is the achievable rate of stream $n$ for user $k$. To indicate the downlink information rate in each stream, by adopting the Shannon capacity equation, $G_{k,n}$ is given as
\begin{equation}
    G_{k,n}(\mathbf{W}_k(t),\mathbf{P}_k(t))) = \log(1 + \gamma_{k,n}(\mathbf{W}_k(t),\mathbf{P}_k(t)))
\end{equation}
where,  for consistency with the notation in the following sections,   $\gamma_{k, n}(\mathbf{W}_k(t),\mathbf{P}_k(t))$ denotes  the Signal-to-Interference-plus-Noise Ratio (SINR)  of stream $n$ for user $k$ as
\begin{equation}
    \gamma_{k, n}(\mathbf{W}_k(t),\mathbf{P}_k(t)) = \frac{\frac{P}{KN_s}\vert \mathbf{w}_{k,n}^{H}(t)\mathbf{H}_k(t)\mathbf{p}_{k,n}(t)\vert^2}{I_{k, n}(t) + I_{\mathrm{c}, k}(t) + \Vert\mathbf{w}_{k,n}(t)\Vert^2\sigma_n^2(t)}\label{SINR}
\end{equation}
where  $\mathbf{W}_k(t) = [\mathbf{w}_{k,1}(t), \dots, \mathbf{w}_{k,N_s}(t)]$ denotes the combining matrix and  $\mathbf{P}_k(t) = [\mathbf{p}_{k,1}(t), \dots, \mathbf{p}_{k,N_s}(t)]$ denotes the precoding matrix. The inter-stream interference for stream $n$ of user $k$  and the multi-user interference for stream $n$ of user $k$  are shown as
\begin{equation}
    I_{k,n}(t) = \sum_{i=1, i \neq n}^{N_s}\frac{P}{KN_s}\vert\mathbf{w}_{k,n}^{H}(t)\mathbf{H}_k(t)\mathbf{p}_{k,i}(t)\vert^2
\end{equation}
and
\begin{equation}
    I_{\mathrm{c}, k}(t) = \sum_{j\in\mathcal{K}, j\neq k}\sum_{i = 1}^{N_s}\frac{P}{KN_s}\vert\mathbf{w}_{k,n}^{H}(t)\mathbf{H}_{k}(t)\mathbf{p}_{j,i}(t)\vert^2
\end{equation}
, respectively.
\subsection{Channel Model}\label{channelmodel}
We assume an extended Saleh-Valenzuela geometric model \cite{raghavan2016beamforming}. The channel between BS and user $k$ is  modeled as a $L$-path channel {as is shown below}
\begin{equation}
    \mathbf{H}_k(t) = \sqrt{\frac{\eta_k MN}{L}}\cdot\sum_{l = 1}^{L}\alpha_{k,l}(t)\cdot\mathbf{u}_{k,l}(t)\mathbf{v}_{k,l}^H(t)
\end{equation}
where $\eta_k$ denotes the large-scale fading coefficient and complex gain $\alpha_{k,l}  (\forall k \in \mathcal{K}, \forall l \in \left\lbrace 1, 2, \ldots, L \right\rbrace
)$ is assumed to remain the same at each time slot and varies between adjacent time slots according to the first-order Gaussian-Markov process
\begin{equation}
    \alpha_{k,l}(t) = \rho \alpha_{k,l}(t-1) + \sqrt{1 - \rho^2}e_{k,l}(t)
\end{equation}
where $e_{k,l}(t) \backsim  \mathcal{CN}(0, 1)$ and $\rho$ is the time correlation coefficient obeying Jakes' model \cite{kim2011mimo}.
\begin{equation}
    \rho = J_o(2\pi f_{\mathrm{d}}\Delta_t\cos{\theta})
\end{equation}
where $f_{\mathrm{d}}$ and $\Delta_t$ denote the Doppler frequency and the channel instantiation interval, respectively, and $J_0$ denotes the first kind $0^{th}$ Bessel function. Since  the users are assumed to move {forward} to the BS or away, i.e., $\theta = 0$ and maximum Doppler frequency $f_{\mathrm{d}}^{\max}$ is achieved which is written as 
\begin{equation}
    \rho = J_o(2\pi f_{\mathrm{d}}^{\max}\Delta_t).
\end{equation}
In the typical case of a uniform linear array (ULA) where the  antennas are deployed at both ends of the {transmission}, the array steering vectors $ \mathbf{u}_{k,l}$ and $\mathbf{v}_{k,l}$ corresponding to the angle of arrival (AoA) $\phi_{\mathrm{A}, k, l}$ and the angle of departure (AoD) $\phi_{\mathrm{D}, k, l}$ in the azimuth are written as
\begin{equation}
    \mathbf{u}_{k,l} = \frac{1}{\sqrt{N}}[1, e^{j2\pi\frac{d}{\lambda}\cos\phi_{\mathrm{A}, k, l}}, \dots, e^{j2\pi\frac{d}{\lambda}(N-1)\cos\phi_{\mathrm{A}, k, l}}]^T
\end{equation}
and
\begin{equation}
    \mathbf{v}_{k,l} = \frac{1}{\sqrt{M}}[1, e^{j2\pi\frac{d}{\lambda}\cos\phi_{\mathrm{D}, k, l}}, \dots, e^{j2\pi\frac{d}{\lambda}(M-1)\cos\phi_{\mathrm{D}, k, l}}]^T
\end{equation}
, respectively, where $\lambda$ is the wavelength of the signal and $d$ denotes the inter-antenna space, which is usually set as $d = \lambda/2$,  $\phi_{\mathrm{A}, k, l}\backsim \mathcal{U}({\theta}_{\mathrm{A}, k, l} - \frac{\delta_{\mathrm{A}}}{2}, {\theta}_{\mathrm{A}, k, l} + \frac{\delta_{\mathrm{A}}}{2})$ and $\phi_{\mathrm{D}, k, l}\backsim \mathcal{U}({\theta}_{\mathrm{D}, k, l} - \frac{\delta_{\mathrm{D}}}{2}, {\theta}_{\mathrm{D}, k, l} + \frac{\delta_{\mathrm{D}}}{2})$ with $\{\theta_{\mathrm{A}, k, l}$, $\theta_{\mathrm{D}, k, l}\}$ referring to the elevation angles and $\{\delta_{\mathrm{A}}$, $\delta_{\mathrm{D}}\}$ denoting the angular spread for arrival and departure, respectively \cite{liang2001downlink}.
\subsection{Problem Formulation}
As described {above}, the system performance heavily relies on precoding and combining vectors design. However, there is an inevitable feedback delay between the time point when the user estimates the channel and  the BS starts transmitting data with the estimated channel fed back by the users. As can be seen in Section \ref{channelmodel}, such delay becomes quite problematic in high mobility scenarios since the channel changes fast and {the} correlation coefficient $\rho$ decreases dramatically. Therefore, it is necessary to develop strategies that {are} robust to feedback
delay and user mobility, which, in this paper, is interpreted as maximizing the sum-rate of $K$ users based on the knowledge of past channels. The problem  can be formulated as follows
\begin{maxi!}
    {\mathbf{W}_k(t),\mathbf{P}_k(t)}{\sum_{k=1}^{K}\sum_{n = 1}^{N_s}G_{k,n}(\mathbf{W}_k(t),\mathbf{P}_k(t))}{\label{pro:original}}{}
    \addConstraint{\Vert\mathbf{p}_{k,n}(t){\Vert}^2 = 1, \forall k,n}\label{p151}
    \addConstraint{\Vert\mathbf{w}_{k,n}(t){\Vert}^2 = 1, \forall k,n}\label{p152}
    \addConstraint{\mathcal{F}(\mathbf{H}_{k}(t^{\prime}) ),\forall k\, \textrm{until}\, t^{\prime} = t-1\, \textrm{are \,available},}\label{p153}
\end{maxi!}
where $\mathcal{F}(\mathbf{H}_{k}(t^{\prime}))$ is a function of $\mathbf{H}_{k}(t^{\prime})$ which is listed in Section \ref{section4}. 
Problem (\ref{pro:original}) aims at optimizing the precoder and combiner to maximize the sum-rate for served users subject to constraints (\ref{p151})- (\ref{p153}), which is  a non-convex problem. To solve this problem,  three efficient DRL-based strategies are proposed in Section  and \ref{section4}.

\section{Multi-agent Deep Reinforcement Learning for Multi-user MIMO Downlink Transmission}\label{section4}

To build up the foundation for the proposed DRL-based designs, an overview of DQN is illustrated first, followed by the description of the state, action, reward function, and three multi-agent DRL-based algorithms for the problem (\ref{pro:original}). 
\subsection{A Brief Overview of  DQN}
\begin{figure}[t]
    \centering
    \includegraphics[width = 8cm]{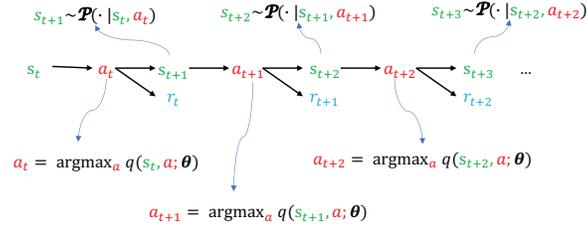}%
    \caption{Markov decision process of  Q-learning.}
   \label{fig:QL}
\end{figure}
In reinforcement learning (RL), an agent learns the optimal action policy to maximize the reward through trial-and-error interactions with the environment.  RL is always formalized as an approach for Makov Decision Process (MDP) problems, which consists of  $\mathcal{S}$, $\mathcal{A}$, $\mathcal{R}$, $\mathcal{P}$, and $\gamma$ referring to a set of states, a set of actions, a reward function, a state transition function, and the discount factor. To be specific, at time $t$, an agent in state $s_t\in \mathcal{S}$ takes an action $a_t \in \mathcal{A}$ according to policy $\pi(a_t|s_t)$, obtains a reward $r_t = \mathcal{R}(a_t, s_t)$ and next state $s_{t+1}\in \mathcal{S}$ with probability $\mathcal{P}(s_t, a_t, s_{t+1})$  in return for the action taken. Formally, each transition (so-called experience of an agent in DQN) can be written as a tuple below
\begin{equation}
    e_t = \langle s_t, a_t, r_t, s_{t+1}\rangle.
\end{equation}
The optimal policy $\pi^*(a_t|s_t)$ is a mapping function between state and action to maximize the future accumulate reward  
\begin{equation}
    R_t = \sum_{\tau = 0}^{\infty}\gamma^{\tau}\mathcal{R}(s_{t+\tau+1}, a_{t+\tau+1})
\end{equation}
where discount factor $\gamma\in [0, 1]$ balances the significance between immediate and future rewards. The optimal policy can be achieved by using dynamic programming (DP) methods that require  {detailed} knowledge of the environment, i.e., $\mathcal{P}(s_t, a_t, s_{t+1})$, which is unavailable due to the  variation of propagation channels.

To tackle this issue, as  illustrated in Fig. \ref{fig:QL}, model-free Q-learning algorithms are demonstrated to continuously improve the policy  through  interactions with the environment. To be specific, the state-action value (called Q-value) is denoted as an expected reward of  $(s, a)$ by policy $\pi$
\begin{equation}
    \mathcal{Q}_{\pi}(s_t, a_t) = \mathbb{E}_{\pi}(R_t|s_t = s, a_t = a)
\end{equation}
where the expectation is calculated over all the possible $(s, a)$ pairs given by policy $\pi$, which can be  iteratively computed  from the Bellman equation
\begin{equation}
    \begin{array}{lr}
    \mathcal{Q}_{\pi}(s_t, a_t) = \mathcal{R}(r_{t+1}|s_t = s, a_t = a) + \gamma\\ \sum_{s^{\prime}\in\mathcal{S}}\bigg(\mathcal{P}(s_{t+1} = s^{\prime}, s_{t} = s, a_{t} =a)\max_{a^{\prime}\in \mathcal{A}}\mathcal{Q}_{\pi}(s^{\prime}, a^{\prime})\bigg)
    \end{array}
\end{equation}
where $\mathcal{P}(s_{t+1} = s^{\prime}, s_{t} = s, a_{t} =a)$ denotes the transition probability from state $s$ to $s^{\prime}$ after taking action $a$.
The optimal policy  returns the maximum expected cumulative reward at each $s$, i.e., $\pi^* = \arg\max_{\pi}\mathcal{Q}^{\pi}(s, a)$. Then the Q-value function can be represented as
\begin{equation}
    \begin{array}{lr}
    \mathcal{Q}_{\pi^*}(s_t, a_t) = r_{t+1}(s_t = s, a_t = a, \pi = \pi^*) \\+ \gamma\sum_{s^{\prime}\in\mathcal{S}}\mathcal{P}(s_{t+1} = s^{\prime}, s_{t} = s, a_{t} =a)\max_{a^{\prime}\in \mathcal{A}}\mathcal{Q}_{\pi^*}(s^{\prime}, a^{\prime}).
    \end{array}
\end{equation}
In classical Q-learning, a Q-value table $q(s, a)$, named as Q-table, is constructed to represent the Q-value function $ \mathcal{Q}_{\pi}(s, a)$. This table consists of a discrete set of $|\mathcal{S}| \times |\mathcal{A}|$ which is randomly initialized. The agent then  {takes} actions according to an $\epsilon$-greedy policy,  receives  reward $r = \mathcal{R}(s, a)$ and transfers to  {the} next state $s_{t+1}$ to complete the experience $e_t$. The Q-table is updated as
\begin{equation}
    q(s_t, a_t)\longleftarrow (1 - \alpha)q(s_t, a_t) + \alpha(r_{t+1} + \gamma\max_{a^{\prime}}q(s_{t+1}, a^{\prime}))
\end{equation}
where $\alpha\in [0, 1)$ is the learning rate. 
However, it is  challenging to directly obtain the optimal $\mathcal{Q}_{\pi^*}(s_t, a_t)$ due to the uncertain variation of the dynamic channel environment, i.e., an unlimited number of states. To address the problems with such an enormous state space, deep Q-network (DQN) is utilized here to approximate the Q-value function, which can be expressed as $q(s_t, a_t, \bm{\theta})$ with $\bm{\theta}$ denoting the weights of DQN. The optimal policy $\pi^*$ can be represented by a group of weights of the DQN. In addition, two techniques are exploited to strengthen the stability of DRL:  target network and experience replay. The target network $q(s_t, a_t, \bm{\bar{\theta}})$ is another network that is initialized with the same set of weights of trained DQN. The target DQN is used to generate the target Q-value which is exploited to formulate the loss function of trained DQN. The weights of target DQN are updated periodically for every fixed number of slots $T_{s}$ by replicating the weights of trained DQN to stabilize the training of trained DQN. The experience replay is intrinsically a first-input-first-output (FIFO) queue that stores  $E_m$ historical experiences in each training slot. During training, $E_b$ experiences are sampled from the experience pool $\mathcal{O}$ to train the trained DQN to minimize the prediction error between the trained DQN and the target DQN. The loss function is defined as
\begin{equation}
    L(\bm{\theta}) = \frac{1}{2E_b}\sum_{\langle s, a, r, s^{\prime}\rangle\in\mathcal{O}}(r^{\prime} - q(s,a;\bm{\theta}))^2
\end{equation}
where $r^{\prime} = r + \gamma\max_{a^{\prime}}q(s^{\prime}, a^{\prime};\bm{\bar{\theta}})$, the weights of DQN $\bm{\theta}$ is updated by adopting a proper optimizer (e.g. RMSprop, Adam, and SGD). The specific gradient update is
\begin{equation}
    \nabla_{\bm{\theta}}L(\bm{\theta}) = \mathbb{E}_{s,a,r,s^{\prime}\in\mathcal{O}}\Big[(r^{\prime} - q(s,a;\bm{\theta})\nabla_{\bm{\theta}}q(s,a;\bm{\theta})\Big]
\end{equation}
\subsection{The Distributed-learning-distributed-processing  DRL-based Algorithm}
In this section,  we cast  the problem (\ref{pro:original}) as a sequential decision-making process and tailor three multi-agent DRL algorithms to solve it. The DRL-based framework is elaborated first, followed by the derived algorithms. To our best knowledge, this is the first paper tackling the problem with 1) receive beamforming with multiple receive antennas, 2) transmit beamforming under imperfect CSIT, 3) multiple streams for each user, and 4) multiple users in a single cell compared with SISO in \cite{nasir2019multi} and MISO in \cite{ge2020deep} with perfect CSIT, respectively. In addition, the PDRL is also firstly demonstrated in this paper to  bridge DDRL and CDRL to balance the performance and complexity.
\subsubsection{Downlink Training and Uplink Feedback}
\begin{figure}[t]
    \centering
    \includegraphics[width = 8cm]{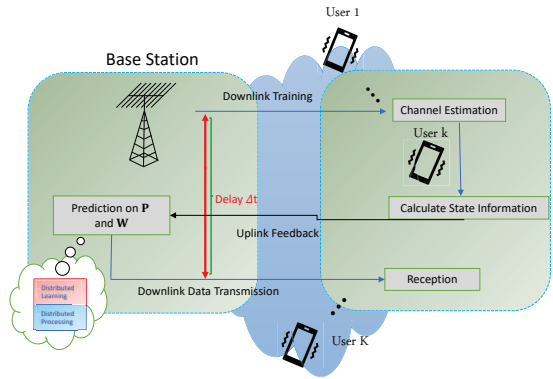}%
    \caption{The downlink training and uplink feedback of proposed DRL framework. The detail structure of the distributed-learning-distributed-processing framework is shown in Fig. \ref{fig:framework}.}
   \label{fig:MADRL}
\end{figure}
\begin{figure}[t]
    \centering
    \includegraphics[width = 8cm]{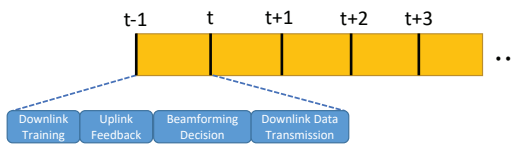}%
    \caption{Timing of time slot $t-1$.}
   \label{fig:PHASES}
\end{figure}

\begin{figure}[t]
    \centering
    \includegraphics[width = 7cm]{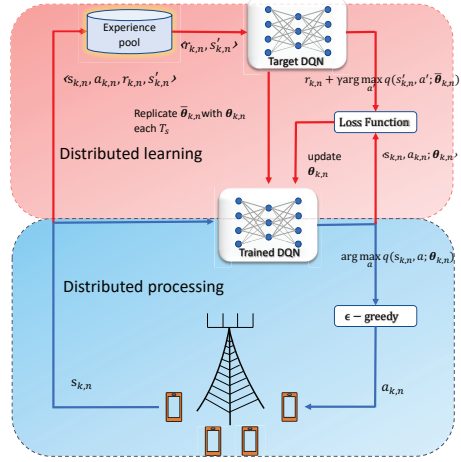}%
    \caption{The framework of distributed-learning-distributed-processing scheme.}
   \label{fig:framework}
\end{figure}
As is shown in Fig. \ref{fig:MADRL} and Fig. \ref{fig:PHASES}, at time slot $t -1$, the BS sends downlink pilots to users, based on which the downlink channels are perfectly estimated. User $k$ can estimate the designed state information in Section \ref{section:state}  and feed it back to the base station.
With feedback from users, the BS can predict the indexes of precoders and combiners for time slot $t$ and start downlink data transmission.
\subsubsection{The Proposed DRL-based Algorithm}
To bring this insight to fruition, each stream is modeled as an agent, totally $KN_s$ agents in our scheme. To be intuitive, we adopt a distributed-learning-distributed-processing framework as shown in Fig. \ref{fig:framework} and demonstrated in Algorithm \ref{alg:DRL_MU_MIMO}.  At the initialization stage, all the $KN_s$ pairs of DQNs are established at the BS. For instance, one pair of  DQNs, namely trained DQN $q(s_{k, n}, a_{k, n}; \bm{\theta}_{k, n})$ and target DQN $q(s_{k, n}, a_{k, n}; \bm{\bar{\theta}}_{k, n})$ is possessed by agent $(k, n)$. The input and output of  trained DQN $q(s_{k, n}, a_{k, n}; \bm{\theta}_{k, n})$ are the local state $s_{k, n}$ and  action $a_{k, n}$. In terms of the distributed learning procedure for agent $(k, n)$, due to the feedback  delay from users, only outdated CSI information is used to formulate the  observations $s_{k, n}$  at the beginning of each time slot. Then, the DRL agent adopts an $\epsilon$-greedy to balance exploitation and exploration by choosing actions,  i.e,  the precoder $\mathbf{p}_{k,n}$, and combiner $\mathbf{w}_{k,n}$ according to $s_{k, n}$, in which the agent executes an action  with probability $\epsilon$ randomly, or executes the action $a_{k, n} = \max_{a}q(s_{k, n}, a;\bm{\theta}_{k, n})$ with probability $1 - \epsilon$. Regarding the distributed learning process, the agent accumulates and stores the experience $e_{k, n} = \langle s_{k, n}, a_{k, n}, r_{k, n}, s_{k, n}^{\prime}\rangle$ into experience pool and the historical experiences  can be utilized to train the DQN with local state-action pairs together with the corresponding reward. Each agent has a profound view of the relationship between local state-action pairs and local long-term reward which, in return, leads the whole system to a distributed-learning-distributed-processing manner. 

\subsubsection{Actions of the proposed multi-agent DRL approach for massive MIMO scenario}
As described in Section \ref{section2}, we aim to optimize the precoder $\mathbf{p}_{k,n}$ and combiner $\mathbf{w}_{k,n}, \forall k, n.$ Then, the problem can be addressed by building two codebooks, i.e. $\mathcal{S}_\mathrm{t}, \mathcal{S}_\mathrm{r}$, which contain $S_{\mathrm{t}}$ and $S_{\mathrm{r}}$ beamforming vectors. In the decision-making stage, each agent chooses one precoder from $\mathcal{S}_\mathrm{t}$ and one combiner from $\mathcal{S}_\mathrm{r}$. The action space can be represented as
\begin{equation}
    \mathcal{A} = \{(\mathbf{c}_\mathrm{t}, \mathbf{c}_\mathrm{r}),\mathbf{c}_\mathrm{t} \in \mathcal{S}_\mathrm{t}, \mathbf{c}_\mathrm{r}\in \mathcal{S}_\mathrm{r}\}
\end{equation}
where $\mathbf{c}_\mathrm{t}$ and $\mathbf{c}_\mathrm{r}$ denote the codewords of two codebooks and the cardinal number of action space  $\mathcal{A}$ is $S_{\mathrm{t}}\times S_{\mathrm{r}}$. The design of codebooks comes from \cite{zou2011beamforming} which is also applied in \cite{ge2020deep, zhang2021joint, chen2021beamforming}, and introduced here as a quantization of  beam directions. To specify each element, we define matrix $\mathbf{C}_\mathrm{t}\in\mathbb{C}^{M\times S_{\mathrm{t}}}$  as
\begin{equation}
    \mathbf{C}_\mathrm{t}[p, q] = \frac{\exp \big(j\frac{2\pi}{T}\lfloor\frac{ M\mathrm{mod} (q + \frac{S_{\mathrm{t}}}{2}, S_{\mathrm{t}})}{S_{\mathrm{t}}/T} \rfloor\big)}{\sqrt{M}}
\end{equation}
where $T$ is the number of available phase values and  $\mathbf{C}_\mathrm{r}\in\mathbb{C}^{N\times S_{\mathrm{r}}}$ can be obtained by substituting the $M$ and $S_{\mathrm{t}}$ with $N$ and $S_{\mathrm{r}}$ accordingly. Each column of $ \mathbf{C}_\mathrm{t}$ and $\mathbf{C}_\mathrm{r}$ corresponds to a specified codeword and the whole matrix forms a beamsteering-based beamformer codebook.
\subsubsection{States of the proposed DRL-based approach for massive MIMO scenarios}\label{section:state}Under the  mobility scenario, the receiver feedback is delayed at time slot $t$,  and the state of agent $(k, n)$ is constructed by the representative feature of observations from the last two successive time slots $t-1$ and $t-2$ without observations from time slot $t$. That is to say, at the beginning of time slot $t - u$, due to the  delay of feedback, the BS is unable to instantaneously obtain the power of the received signal, i.e., $|\mathbf{w}^H_{k,n}(t)\mathbf{H}_k(t)\mathbf{p}_{k,n}(t)|^2$ and $|\mathbf{w}^H_{k,n}(t)\mathbf{H}_k(t-1)\mathbf{p}_{k,n}(t)|^2$. However, the historical feedback, i.e.,  $|\mathbf{w}^H_{k,n}(t-1)\mathbf{H}_k(t-1)\mathbf{p}_{k,n}(t-1)|^2$ and  $|\mathbf{w}^H_{k,n}(t-1)\mathbf{H}_k(t-2)\mathbf{p}_{k,n}(t-1)|^2$ are usually available to the BS. Based on this assumption, the state $s_{k,n}(t)$ is designed as follows
\begin{itemize}
    \item The "desired" information of  the agent $(k, n)$ which consists of  5 parameters, i.e., the channel gain $|\mathbf{w}^H_{k,n}(t-1)\mathbf{H}_k(t-1)\mathbf{p}_{k,n}(t-1)|^2$, the chosen
index of precoder $U_{k,n}(t-1)$,  the chosen
index of combiner $V_{k,n}(t-1)$, the achievable rate of stream $n$ for user $k$, i.e., $G_{k,n}(\mathbf{P}_k(t-1), \mathbf{W}_k(t-1))$, and the interference-plus-noise $I_{k, n}(t-1) + I_{\mathrm{c}, k}(t-1) + \sigma_k^2$.
    \item Interference information of the agent $(k, n)$ which is represented by 8 parameters, i.e., $\{\sum_{i = 1, i\neq n}^{N_s}|\mathbf{w}^H_{k,n}(t-u)\mathbf{H}_k(t-u)\mathbf{p}_{k,i}(t-u)|^2, \sum_{i = 1, i\neq n}^{N_s}|\mathbf{w}^H_{k,n}(t-1 - u)\mathbf{H}_k(t-u)\mathbf{p}_{k,i}(t-1-u)|^2, \sum_{j\neq k}\sum_{i = 1}^{N_s}|\mathbf{w}^H_{k,n}(t-u)\mathbf{H}_k(t-u)\mathbf{p}_{j,i}(t-u)|^2,
    \sum_{j\neq k}\sum_{i = 1}^{N_s}|\mathbf{w}^H_{k,n}(t-1- u)\mathbf{H}_k(t-u)\mathbf{p}_{j,i}(t-1-u)|^2 \lvert u\in\{1, 2\}\}$. It is worth noting here that in such a system, the interference information plays a key role in  the maximization of its own information rate (the rate of stream $n$ of user $k$), which, thus, should be included in state space.
    \item The information of  agent $(j, i), (j, i)\neq (k, n), \forall j, i$  consists of of $10(KN-1)$ terms, i.e.,$\{U_{j,i}(t-u), V_{j,i}(t-u), G_{j,i}(\mathbf{M}_j(t-u), \frac{P}{NK}\vert\mathbf{w}^H_{j,i}(t-u)\mathbf{H}_j(t-u)\mathbf{p}_{j,i}(t-u)\vert^2, \frac{P}{NK}\vert\mathbf{w}^H_{j,i}(t-u)\mathbf{H}_j(t-u)\mathbf{p}_{k,n}(t-u)\vert^2\lvert u\in\{1, 2\}\}$. The information of other agents plays an irreplaceable role for agent $k$ to minimize the interference it causes to them, which, thus,  should be included in state space.

\end{itemize}
To sum up, the cardinal number of state space is $10KN_s + 3$. Note that the adopted design is not guaranteed to be the optimal one but empirically achieves a good performance as demonstrated with evaluation results in Section \ref{section4}. The output size of the DQN is $S = S_{\mathrm{t}}S_{\mathrm{r}}$ which is equal to the number of available actions. 
\subsubsection{The reward of the proposed DRL-based approach for the massive MIMO scenario} In this massive MIMO scenario, if agent $(k, n)$ only tries to maximize the achievable rate of the stream $(k, n)$ without taking the inter-stream and multi-user interference into consideration,  a large interference will be delivered to other agents. Therefore, our proposed reward function $r_{k,n}$ consists of penalty coefficient $\lambda$ and penalty term $P_{k,n}(\mathbf{W}_k(t), \mathbf{P}_k(t))$ to quantify the adverse impact each agent causes to other agents. The penalty term $P_{k,n}(\mathbf{W}_k(t), \mathbf{P}_k(t))$ is given as 
\begin{equation}
    \begin{array}{lr}    
   P_{k,n}(\mathbf{W}_k(t), \mathbf{P}_k(t)) \\= \sum_{j = 1, j\neq k}^{K}\sum_{i = 1 }^{N_s}\bigg( \log_2(1 + \frac{\frac{P}{KN_s}|\mathbf{w}^H_{j,i}(t)\mathbf{H}_j(t)\mathbf{p}_{j,i}(t)|^2}{\sigma^2 + \hat{I}_{k, n_1}(t) + \hat{I}_{\mathrm{c}_1, k}(t)})  - G_{j,i}(\mathbf{W}_k(t), \mathbf{P}_k(t))\bigg) \\ + \sum_{i = 1, i\neq n }^{N_s}\bigg( \log_2(1 + \frac{\frac{P}{KN_s}|\mathbf{w}^H_{k,i}(t)\mathbf{H}_k(t)\mathbf{p}_{k,i}(t)|^2}{\sigma^2 + \hat{I}_{k, n_2}(t) + \hat{I}_{\mathrm{c}_2, k}(t)}) - G_{k,i}(\mathbf{W}_k(t), \mathbf{P}_k(t))\bigg)
   \end{array}
\end{equation}
where $\hat{I}_{k, n_1}(t)$, $\hat{I}_{k, n_2}(t)$, $\hat{I}_{\mathrm{c}_1, k}(t)$ and $\hat{I}_{\mathrm{c}_2, k}(t)$ are given by
\begin{equation}
    \hat{I}_{k, n_1}(t) = \sum_{i = 1, i \neq l}^{N_s}\frac{P}{KN_s}\vert\mathbf{w}^H_{j,i}(t)\mathbf{H}_j(t)\mathbf{p}_{k,i}(t)\vert^2,
\end{equation}
\begin{equation}
    \begin{array}{lr}
    \hat{I}_{\mathrm{c}_1, k}(t) = \sum_{q\in\mathcal{K}, q \neq k}\sum_{i = 1}^{N_s}
    \frac{P}{KN_s}\vert\mathbf{w}^H_{j,i}(t)\mathbf{H}_{j}(t)\mathbf{p}_{q,i}(t)\vert^2
    -\frac{P}{KN_s}
    \vert\mathbf{w}^H_{j,i}(t)\mathbf{H}_{j}(t)\mathbf{p}_{j,i}(t)\vert^2,
    \end{array}
\end{equation}
\begin{equation}
    \hat{I}_{k, n_2}(t) = \sum_{h = 1, h \neq i, n}^{N_s}\frac{P}{KN_s}\vert\mathbf{w}^H_{k,i}(t)\mathbf{H}_k(t)\mathbf{p}_{k,h}(t)\vert^2
\end{equation}
, and
\begin{equation}
    \hat{I}_{\mathrm{c}_2, k}(t) = \sum_{q\in\mathcal{K}, q \neq k}\sum_{h = 1}^{N_s}\frac{P}{KN_s}\vert\mathbf{w}^H_{k,i}(t)\mathbf{H}_{k}(t)\mathbf{p}_{q,h}(t)\vert^2 
\end{equation}
, respectively.
Note that $P_{k,n}(\mathbf{W}_k(t), \mathbf{P}_k(t))$ is always a positive value due to the extraction of the interference from a specified stream.
Then, the achievable rate for stream $(k,n)$, i.e., $G_{k,n}(\mathbf{W}_k(t), \mathbf{P}_k(t))$, is added into $r_{k, n}$ to highlight the contribution of agent $(k, n)$ to the total information rate. Hence,  $r_{k, n}$ at time slot $t$ is given as 
\begin{equation}
    r_{k,n}(t) = G_{k,n}(\mathbf{W}_k(t), \mathbf{P}_k(t)) - \lambda P_{k,n}(\mathbf{W}_k(t), \mathbf{P}_k(t))
\end{equation}
where  penalty coefficient $\lambda$ is used here as a weight parameter to manipulate the amount of negative effect in the reward function.
In regard to  the reward function, the rationale behind such a design is to maximize the achievable rate of improvement if the interference caused by stream $(k, n)$ is totally eliminated. This design not only maximizes the achievable rate of stream $(k, n)$, i.e., $G_{k,n}$ but also minimizes the negative effect it causes to other streams, i.e., $P_{k,n}$.   Similar designs are comprehensively discussed in \cite{nasir2019multi, ge2020deep} which also confirm that a well-formulated reward function should act as a catalyst of the best decisions obtained by multiple agents.

\begin{table*}[ht] \caption{comparison between strategies} 
\centering      
\begin{adjustwidth}{0cm}{}
\begin{tabular}{c c c c}
\hline\hline                        
& FDD $\&$ ZF-CI\cite{bjornson2016massive} & TDD $\&$ ZF-CI\cite{bjornson2016massive} & FDD $\&$ MA-DRL  \\ [0.5ex] 
\hline   {\makecell[c]{CSI Overhead \\Uplink}}& {\makecell[c]{ $MN + KN$\\($MN$ channel coefficients \\and $KN$ CSI pilot symbols) }}& $KN$ sounding pilot symbols &{\makecell[c]{  $MN + KN + 11$(\\$MN$ channel coefficients, \\ $KN$ CSI pilot symbols,\\ and $\{|\mathbf{w}^H_{k,n}(t-1)\mathbf{H}_k(t-1)\mathbf{p}_{k,n}(t-1)|^2$, \\$G_{k,n}(\mathbf{P}_k(t-1), \mathbf{W}_k(t-1))$,\\ $I_{k, n}(t-1) + I_{\mathrm{c}, k}(t-1) + \sigma_k^2$,\\ $\sum_{i = 1, i\neq n}^{N_s}|\mathbf{w}^H_{k,n}(t-v- u)\mathbf{H}_k(t-u)\mathbf{p}_{k,i}(t-v-u)|^2,$\\ $\sum_{j\neq k}\sum_{i = 1}^{N_s}|\mathbf{w}^H_{k,n}(t-v-u)\mathbf{H}_k(t-u)\mathbf{p}_{j,i}(t-v-u)|^2$,\\$u\in\{1,2\},v\in\{0,1\}$\}}}\\ \hline {\makecell[c]{ CSI Overhead\\ Downlink}} & $MN$ CSI pilot symbols& 0 &{\makecell[c]{ $MN + 2N$\\ ($MN$ channel coefficients and indexes\\ of precoders and combiner $\mathbf{p}_{k,n}$ and $\mathbf{w}_{k,n}$) }} \\ \hline {\makecell[c]{Computing Complexity\\ of Precoding and \\ Combining Matrices}} & $\mathcal{O}((MN)^3)$ & $\mathcal{O}((MN)^3)$  & $\mathcal{O}((10KN + 3)L_1 + L_1L_2 + L_2S)$  \\ [1ex] \hline   
\end{tabular} \label{table:complexity}  
\end{adjustwidth}
\end{table*} 
\subsubsection{Discussion on the overhead and complexity of the proposed framework} As is shown in Table \ref{table:complexity}, if the base station has to tell the users what combiner to use, then it can consume additional overhead on the downlink transmission. Fortunately, this overhead is negligible since only the indexes of combiners are delivered to users. Note that the precoders are also sent to terminals for the calculation of state information listed in the table. {This reduces the computation burden on the base station for processing this state information.}

In terms of the computational complexity of precoders and combiners in the demonstrated DRL-based  {approaches}, the designed structure of target/trained DQNs includes four fully connected layers. {Specifically, the input layer consists of $10KNs + 3$ neurons, followed by two hidden layers with $L_1$ and $L_2$ neurons and a specified activation function. The fourth layer serves as the output layer with $S$ neurons. We employ two hidden layers in our design, as a two-layer feedforward neural network is sufficient to approximate any nonlinear continuous function based on the $\textit{universal approximation theorem}$ \cite{lu2017expressive}. The computational complexity of fully connected DNN  can be written as $\mathcal{O}((10KN_s + 3)L_1 + L_1L_2 + L_2S)$ for each agent. This is much smaller than that of ZF-CI scheme due to the fact that ZF-CI  involves matrix inversion which limits the scalability to a large number of transmit and receive antennas}.
\begin{remark}
\textit{Note that different from \cite{kim2020massive} where the mobility estimation and channel prediction are needed, our work does not predict the channels sequentially. In this paper, we demonstrated a low complexity and efficient DRL-based framework and as this is the first work proposing DRL-based joint transmit and receiver beamforming for massive MIMO downlink transmission, we would like to keep the benchmarks as clear and simple as possible such that researchers can understand the fundamental benefits of the proposed strategies and carry on their studies in more practical
scenarios in the future. The comparison with mobility estimation and channel prediction methods (such as VFK and MLP  methods in \cite{kim2020massive})  could be addressed in future research, but  not the scope of this paper.}
\end{remark}

\begin{algorithm}[t]
    \caption{DDRL  Algorithm}\label{alg:DRL_MU_MIMO}
    \begin{algorithmic}[1]
        \State{$\textbf{Initialize}$: Establish a trained DQN and target DQN with random weights $\bm{\theta}_{k,n}$ and $\bm{\bar{\theta}}_{k,n}$, respectively, $\forall k \in \left\lbrace 1, 2, \ldots, K \right\rbrace, \forall n \in \left\lbrace 1, 2, \ldots, N_s \right\rbrace$, update the weights of $\bm{\bar{\theta}}_{k,n}$ with $\bm{\theta}_{k,n}$.}
        
        \State{In the first $E_s$ time slots,  agent $(k, n)$ randomly selects an action from action space $\mathcal{A}$, and stores the corresponding experience $\langle s_{k,n}, a_{k,n}, r_{k,n}, s_{k,n}^{\prime}\rangle$ in its pool, $\forall k,n$.}
        
        \State{$\textbf{for}$  each time slot $t$ $\textbf{do}$}
             \State{\quad$\textbf{for}$ each agent $(k, n)$ $\textbf{do}$}
            \State{\quad\quad Obtain state $s_{k, n}$ from the observation of agent $(k, n)$.}   
            \State{\quad\quad Generate a random number $\omega$.}
            \State{\quad\quad\textbf{If} $\omega < \epsilon$ $\textbf{then}$:}
            \State{\quad\quad\quad Randomly select an action in action space $\mathcal{A}$.}
            \State{$\quad\quad\textbf{Else}$}
            \State{\quad\quad\quad Choose the action $a_{k,n}$ according to the Q-function $q(s_{k,n}. a; \bm{\theta}_{k,n}), \forall k, n$ }
            \State{\quad\quad\textbf{End if }.}
        \State{\quad\quad Agent $(k, n)$ executes the  $a_{k,n}$, immediately receives the reward $r_{k, n}$ and steps into next state $s_{k, n}^{\prime}, \forall k, n$.}
        \State{\quad\quad Agent $(k, n)$ puts experience $\langle s_{k,n}, a_{k,n}, r_{k,n}, s_{k,n}^{\prime}\rangle$ into experience pool $\mathcal{O}_{k, n}$, randomly samples a minibatch with size $E_b$. Then, the weights of trained DQN $\bm{\theta}_{k, n}$ are updated using back propagation approach. The weights of target DQN $\bm{\bar{\theta}}_{k, n}$ is updated every $T_s$ steps.}
        \State{\quad\textbf{\textbf{end for}}}
        \State{\textbf{\textbf{end for}}}
    \end{algorithmic}
\end{algorithm}
\subsection{The Low-complexity Centralized-learning-distributed-processing  DRL-based Algorithm}
In this section, we demonstrate an extra algorithm for the problem (\ref{pro:original}) for three reasons. \textit{First}, a lower computation complexity is achieved in the centralized scheme by  building and training on an extra pair of DQNs  instead of distributedly training  with $KN_s$ agents. \textit{Second}, a lower storage space is required with only a central experience pool during the learning process. \textit{Third}, by saving and sampling the experiences from all distributed agents, the central agent can learn the common features from the channels of all users and intelligently guide the decision-making procedure of all distributed agents. {Things need to be noted that CDRL is trained more efficiently using parameter sharing, 
 which is based on homogeneous agents. This allows the policy to be trained with the experiences of all agents simultaneously. However, it still allows different actions between agents due to the fact that each agent receives different observations. This algorithm focuses on the decentralized parameter-sharing training scheme since we found it to be scalable if we continue to increase the number of users and streams.} 

There are also some similarities between CDRL and DDRL. On the one hand, they have the same  state, action, and reward function without the necessity of designing new ones. On the other hand, the executing phase is also performed by distributed agents. 

The whole process is shown in Algorithm \ref{alg:CDRL_MU_MIMO}. At the initialization stage, only one pair of target and trained DQNs is built for the central agent. For each distributed agent,  one trained DQN is established. In the first several time slots, each agent randomly selects an action and saves the experiences into the central experience  pool. When the episode begins, the central agent adopts an $\epsilon$-greedy strategy to balance exploitation and exploration  so as to find the optimal policy. After learning from the sample experiences, the central agent broadcasts the updated weights of the central trained DQN to all other distributed agents for decision-making purposes.

\begin{algorithm}[t]
    \caption{CDRL  Algorithm}\label{alg:CDRL_MU_MIMO}
    \begin{algorithmic}[1]
        \State{$\textbf{Initialize}$: Establish a central trained DQN and central target DQN with random weights $\bm{\theta}_{\mathrm{c}}$ and $\bm{\bar{\theta}}_{\mathrm{c}}$ for the central agent, update the weights of $\bm{\bar{\theta}}_{\mathrm{c}}$ with $\bm{\theta}_{\mathrm{c}}$.
        Establish a  trained DQN with random weight $\bm{\theta}_{k,n}$, $\forall k \in \left\lbrace 1, 2, \ldots, K \right\rbrace, \forall n \in \left\lbrace 1, 2, \ldots, N_s \right\rbrace$ for each distributed agent.}
        
        \State{In the first $E_s$ time slots,  agent $(k, n)$ randomly selects an action from action space $\mathcal{A}$, and stores the  experience $\langle s_{k,n}, a_{k,n}, r_{k,n}, s_{k,n}^{\prime}\rangle, \forall k, n$ in  the experience pool of central agent  $\mathcal{O}_{\mathrm{c}}$.}
        
        \State{$\textbf{for}$  each time slot $t$ $\textbf{do}$}
             \State{\quad$\textbf{for}$ each agent $(k, n)$ $\textbf{do}$}
            \State{\quad\quad Obtain state $s_{k, n}$ from the observation of agent $(k, n)$.}   
            \State{\quad\quad Generate a random number $\omega$.}
            \State{\quad\quad\textbf{If} $\omega < \epsilon$ $\textbf{then}$:}
            \State{\quad\quad\quad Randomly select an action in action space $\mathcal{A}$.}
            \State{$\quad\quad\textbf{Else}$}
            \State{\quad\quad\quad Choose the action $a_{k,n}$ according to the Q-function $q(s_{k,n}. a; \bm{\theta}_{k,n}), \forall k, n$ }
            \State{\quad\quad\textbf{End if }.}
        \State{\quad\quad Agent $(k, n)$ executes the  $a_{k,n}$, immediately receives the reward $r_{k, n}$ and steps into next state $s_{k, n}^{\prime}, \forall k, n$.}
        \State{\quad\quad Agent $(k, n)$ puts experience $\langle s_{k,n}, a_{k,n}, r_{k,n}, s_{k,n}^{\prime}\rangle$ into central experience pool $\mathcal{O}_{\mathrm{c}}$.}
        \State{\quad\textbf{\textbf{end for}}}
        \State{\quad Central agent  randomly samples a minibatch with size $E_b$. Then, the weights of central trained DQN $\bm{\theta}_{\mathrm{c}}$ are updated using the back propagation approach. The weights of target DQN $\bm{\bar{\theta}}_{\mathrm{c}}$ is updated every $T_s$ steps. Then, central agent broadcasts the weights $\bm{\theta}_{\mathrm{c}}$ to all the distributed agents, i.e., $\bm{\theta}_{k,n} = \bm{\theta}_{\mathrm{c}}, \forall k, n.$}
        
        \State{\textbf{\textbf{end for}}}
    \end{algorithmic}
\end{algorithm}
\subsection{Bridging the DDRL and CDRL: Partial-distributed-learning-distributed-processing Scheme}
In contrast with DDRL and CDRL, the partial-distributed-learning-distributed-processing DRL-based scheme (PDRL) offers a more flexible solution to the problem (\ref{pro:original}) by modeling each user as an agent. In the extreme case of $N_s = 1, K > 1$, PDRL boils down to DDRL by simply treating each stream as an agent. In the other extreme case of $K = 1, N_s > 1$,  PDRL boils down to CDRL by forcing one central agent to do the training work. Compared with CDRL, PDRL demonstrates better performance-complexity balance by  learning  the representative features of the propagation environment for a specified user which is demonstrated in Fig. \ref{fig:partial}. The whole algorithm is illustrated in Algorithm \ref{alg:PDRL_MU_MIMO}.
\section{Result Evaluation}\label{Section5}
This section demonstrates the performance of our proposed multi-agent DRL-based algorithm to maximize the average throughput of all the users. We first illustrate the simulation setup, followed by the simulation results in different scenarios. 
\subsection{Simulation Setup}
We consider a downlink transmission from one BS to multiple users. The BS serves $K = 4$ users in a single cell. The maximum {transmit} power $P$ is fixed to {20 dBm} and noise variance $\sigma^2$ at users  is fixed to {-114 dBm}. The BS is  equipped with $M = 32$ transmit antennas and the users are equipped with  $N = 4$ receive antennas unless otherwise stated.
Without loss of generality, the uniform linear array (ULA) is equipped in both transmitter and receiver sides with half-wavelength inter-antenna spacing. 
The large-scale channel fading is characterized by the log-distance path-loss model 
expressed below
\begin{equation}
     \eta =   L(d_0) + 10\omega\log_{10}\frac{d}{d_0}.
\end{equation}
where {$d = 10$ m}  is the BS-user distance. According to Table III of \cite{smulders2009statistical}, the value of $L(d_0) $ for {$d_0 = 1$ m} is {68 dB} and fading coefficient $\omega$ is 1.7. In terms of the shadowing model, the log-normal shadowing standard deviation $\beta_k$ is set to {1.8 dB}.
The small-scale fading channel is generated according to the channel model introduced in Section \ref{section2}.
Regarding the parameters of Jake's model with user speed {3.55 km/h}, the maximum Doppler frequency $f_{\mathrm{d}}^{\max}$ and  channel instantiation interval $T_i$ are set as {800 Hz} and {$1\times10^{-3}$ s}, respectively\cite{smulders2009statistical}. The corresponding correlation coefficient $\rho$ is 0.6514$\approx$ 0.65.  

As is illustrated in Fig. \ref{fig:framework}, the whole framework can be divided into 2 phases, the  learning phase, and the  processing phase. Before {the} learning phase, we randomly generate channels obeying Jake's model, randomly choose actions, observe the reward, and accumulate and store the corresponding experiences into the experience pool with size 1000 for the first 200 time slots, i.e., $E_m = 1000, E_s = 200$. In addition, the mini-batch size $E_b$ is set as 32. Stepping into the learning stage, for the DNN, the number of neurons in two hidden layers, i.e, $L_1, L_2$, are both set as 256, followed by the $ReLu$ activation function. The initial learning rate $\alpha(0)$ is 5$e^{-3}$ and the decaying rate $d_c$ is $10^{-4}$ such that the learning rate continues to decay with {the} number of time slots following $\alpha(t) = \alpha(t-1) * \frac{1}{1 + d_ct}$. In terms of optimization, the adaptive moment estimation ($Adam$) is utilized  to prevent the diminishing learning rate problem. To minimize the prediction error between trained DQN and target DQN, the weights of trained DQN are substituted into target DQN every 120 time slots, i.e., $T_s = 120$ with discount factor $\gamma$ and penalty coefficient $\lambda$ set as 0.1 and 1, respectively. During the processing phase, for $\epsilon$-greedy strategy, we set the initial exploration coefficient $\epsilon$ as $0.7$ which decays exponentially to 0.001. Note that the adopted parameters are not guaranteed to be optimal ones, which   experimentally  perform well in this setup. In the legend of simulation figures, DDRL and CDRL come from Algorithm \ref{alg:DRL_MU_MIMO} and Algorithm \ref{alg:CDRL_MU_MIMO}, respectively. {The value} of each point is a moving average over the previous 500 time slots unless otherwise stated.

To demonstrate the effectiveness of our DRL-based approaches, four benchmark schemes are evaluated, which are as follows:
\begin{itemize}
\item ZF-CI PCSI: Each agent executes the action from the scheme in \cite{sung2009generalized} with instantaneous and perfect CSI, i.e., $\mathbf{H}_{k}(t), \forall k$.
\item SAH: This approach stores the most recent estimated channel, i.e., $\mathbf{H}_{k}(t-1), \forall k$ and this approach always sends the channel coefficients to the base station, which will be used for  calculating the precoders  using ZF-CI. This  strategy   essentially  ignores the non-negligible  delay between   the channel estimation   and the time point when the actual DL transmission happens \cite{zhang2022predicting}. When $\rho = 1$, SAH is the same as ZF-CI PCSI. SAH only captures delay but assumes perfect knowledge of CSI at $t-1$.
\item Random: Each agent randomly chooses actions. The performance serves as a {lower bound} in the simulation.
\item  Greedy Beam Selection {(GBS)}: Each agent exhaustively selects an action in a greedy manner, the actions with the highest sum information rate are chosen as the solution for each channel realization. The benchmark serves as the {upper bound} for DRL-based strategies. Note that the size of the beam selection set increases exponentially with the size of the codebooks (${(NK)}^{S_tS_r}$). For instance, when $K = 4, N= 1, S_\mathrm{t} = 32, S_\mathrm{r} = 4$,  the total number of action combination is $4^{32}$ which is quite large considering the hardware constraint. Thus, we consider $(8, 1)$ in this benchmark.
\end{itemize}

\subsection{DDRL vs CDRL}

\begin{figure}[t]
  \centering
    
    \begin{minipage}[b]{0.45\textwidth} 
    \centering 
    \includegraphics[scale = 0.5]{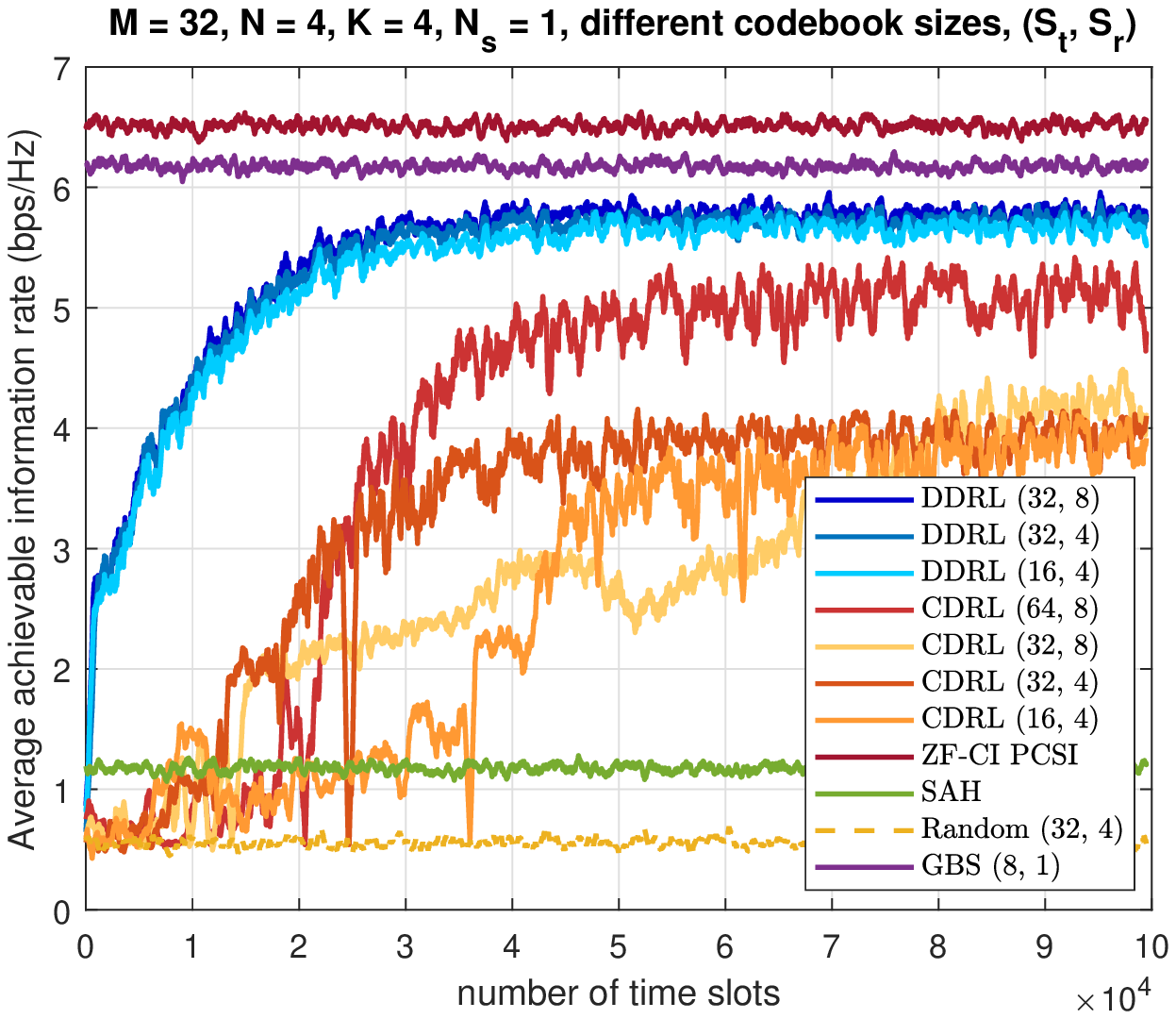}%
    
  \caption{Average information rate versus the number of time slots with different codebook sizes $(S_{\mathrm{t}}, S_{\mathrm{r}})$. }
    \label{fig:cbsize1stream}
    \end{minipage}
    \begin{minipage}[b]{0.45\textwidth} 
    \centering 
    \hspace{-7mm}
    
  \includegraphics[scale = 0.5]{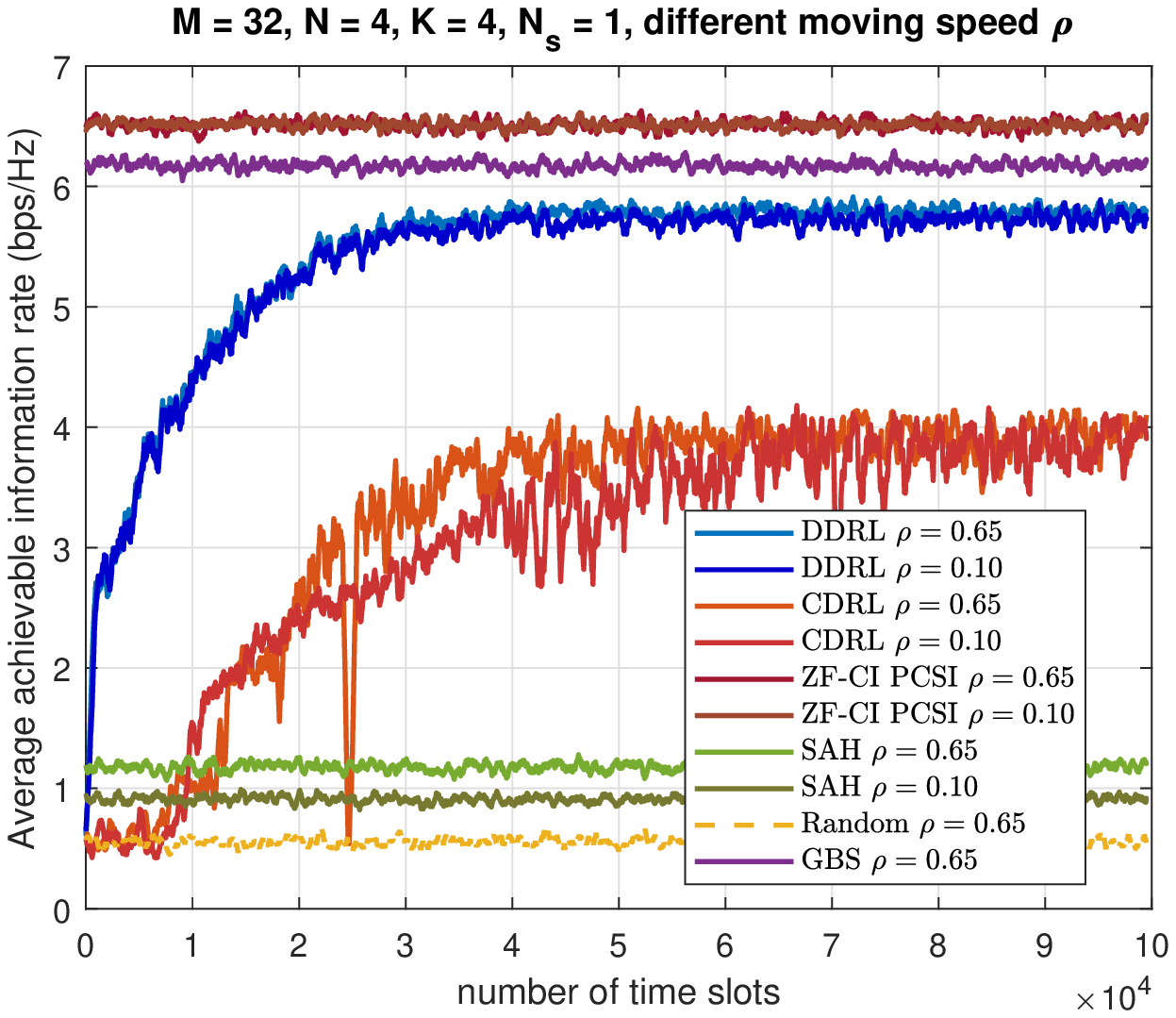}%
    
    \caption{Average achievable information rate versus the number of time slots with different correlation coefficients. }
   \label{fig:speed1stream}
   \end{minipage}
   
 \end{figure}

\begin{figure}[t]
  \centering
    \begin{minipage}[b]{0.45\textwidth} 
    \centering 
    \includegraphics[scale = 0.5]{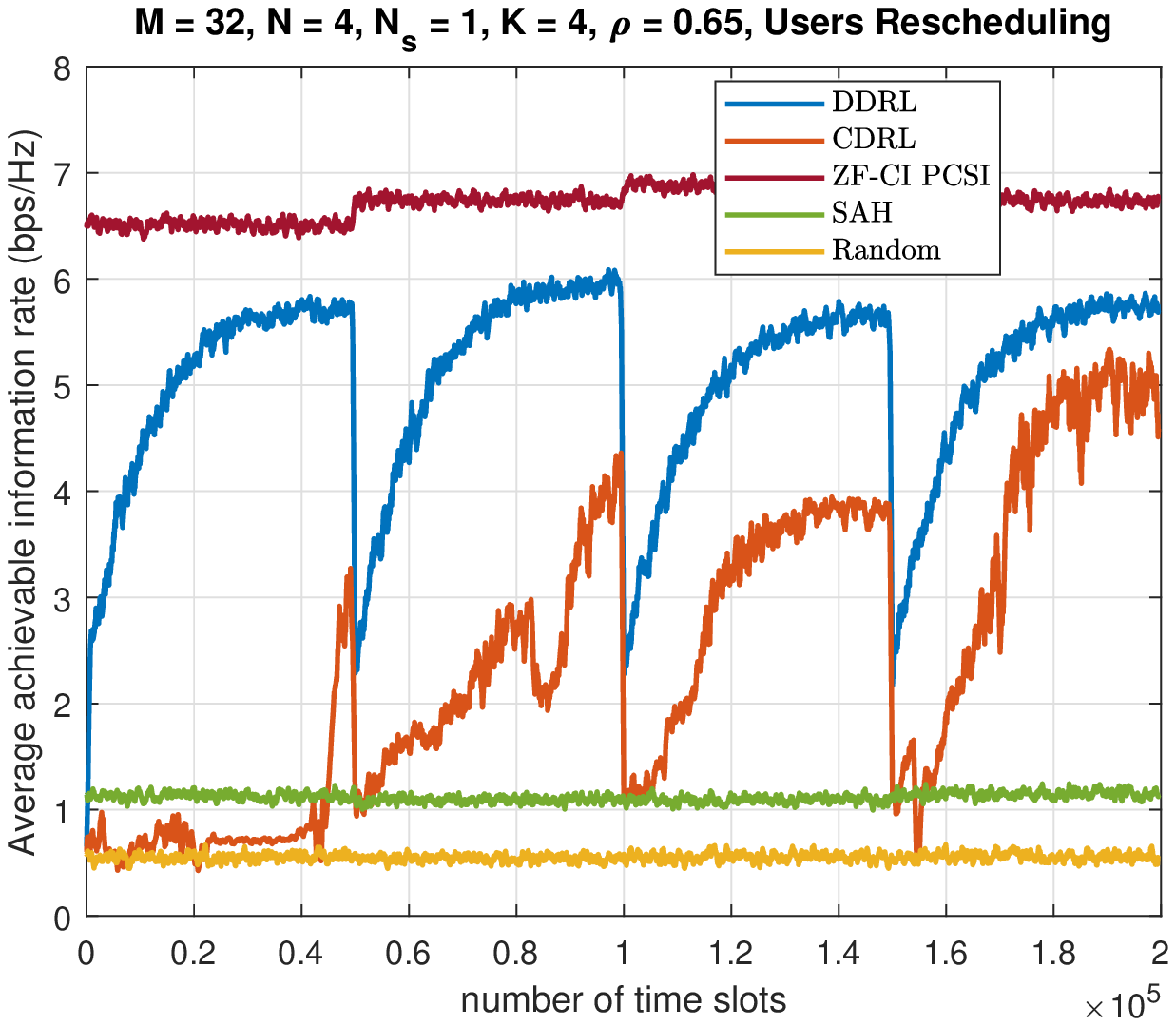}%

  \caption{Average information rate versus number of time slots with users rescheduling  at 5e4th, 1e5th and 15e4th time slot.}
    \label{fig:rescheduling}
    \end{minipage}
    \begin{minipage}[b]{0.45\textwidth} 
    \centering 
    \hspace{-10mm}
    
  \includegraphics[scale = 0.51]{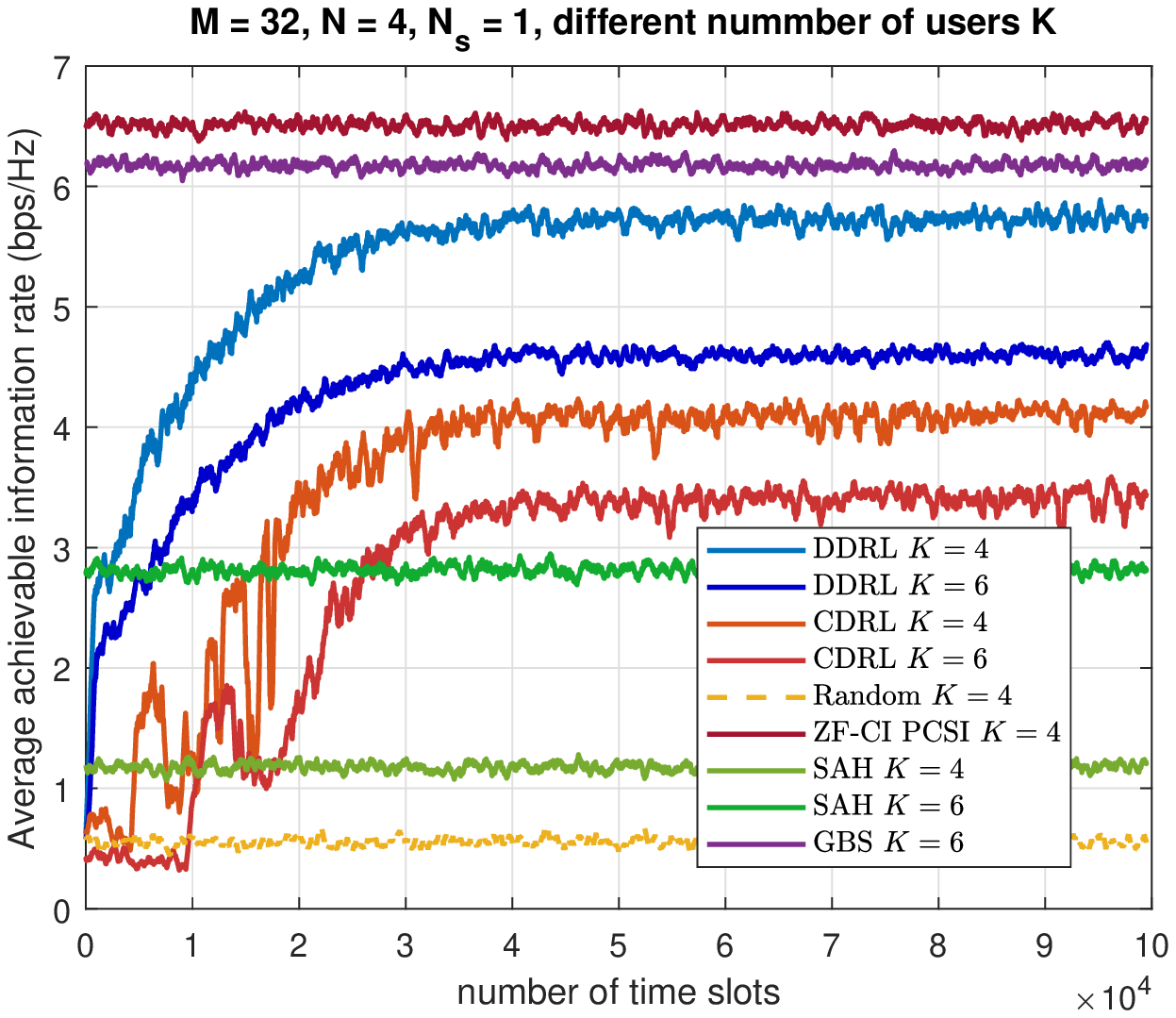}%
    
    \caption{Average information rate versus number of time slots with different number of users $K$.}
   \label{fig:users}
   \end{minipage}
   
 \end{figure}
Fig. \ref{fig:cbsize1stream} depicts the average achievable information rate versus the number of time slots with different numbers of transmit and receive beamformer codebook size $(S_{\mathrm{t}}, S_{\mathrm{r}})$. A \textit{first} observation is that  the performance gaps between two DRL-based schemes and SAH are gradually increased with the number of $S_{\mathrm{t}}$ and $S_{\mathrm{r}}$ and DDRL  roughly observes a gain of 380$\%$ over SAH when $S_{\mathrm{t}} = 32$ and $S_{\mathrm{r}} = 4$. The reason behind such a phenomenon is that,  in  DRL-based strategies, better interference management can be achieved by higher resolution in the codebook which significantly reduces the quantization error and effectively alleviates the interference from other streams.{ A \textit{second} observation is that  
DDRL  can achieve nearly 90$\%$ of the system capacity of the ZF-CI PCSI and $95\%$ of the Beam Selection strategy, utilizing only  a few  pieces of information in the designed features from itself and other users. An interpretation is that the lack of instantaneous CSI and imperfect codebook design degrade the system performance and this 10$\%$ and 5$\%$gap cannot be fully  eliminated A \textit{third} observation is that  DDRL $(32, 8)$ only demonstrates slightly better performance than DDRL $(32, 4)$ and DDRL $(16, 4)$  which shows of robustness on codebook size.} A \textit{fourth} observation is that the CDRL always demonstrates instability before  convergence. An explanation is that the huge differences between the dynamic environment of different users make it extremely difficult to find the commonality among them. Then, each agent could be misled by the experiences of other agents, which, thus, results in fluctuations before convergence and degradation in system performance. Conversely, in DDRL, each agent selects a specific precoder and combiner for its intended stream which is relevant to its propagation environment and is considerably different among streams. This local adaptability greatly improves the performance of DDRL. 
After comprehensive considerations among  computation complexity, system performance, and convergence speed, $(32, 4)$ is chosen as a codebook baseline in the simulations of both DRL-based schemes. Compared with the Greedy Beam Selection method, the DRL-based strategies reveal extremely lower complexity on large action space but achieve roughly 95$\%$  of Greedy Beam Selection performance.\ref{table:time}. {Hence, CDRL is not suitable for practical systems where a large codebook is not available while DDRL demonstrates more robustness regarding codebook size but requires much more amount of memory and computing resources for training.}  We implemented the demonstrated algorithms with TensorFlow in a general computer, i.e., i7-8700 CPU, 3.20 GHz. The running time for different algorithms is listed in {Table. \ref{table:time}}.  

\begin{table*}[ht] 
\centering      
\caption{Running Time For Each Channel Realizarion} 
\begin{tabular}{ |c|c|c|c|c|c| }  
\hline\hline                      
 & DDRL (32, 4) &  ZF PCSI  & SAH & Beam Selection (8, 1) & Random Selection\\  
\hline   
Time& 0.2s  & 0.8s &  0.8s & 10s & 0.1s \\
\hline  
\end{tabular} \label{table:time}  

\end{table*} 

Fig. \ref{fig:speed1stream} exhibits the average achievable information rate versus the number of time slots with different values of correlation coefficient $\rho$. The DDRL scheme with $\rho = 0.65$ and $\rho = 0.1$
can exceed the benchmark SAH with approximately 380$\%$ and  500$\%$, respectively. This result greatly embodies the superiority of our DRL-based framework over the traditional  massive MIMO optimizing scheme in mobility scenarios since a 20$\%$ performance degradation is caused by the fast-changing channels in SAH. In addition, it can  be observed that  DDRL with $\rho = 0.1$ demonstrates a slightly lower performance than $\rho = 0.65$ which is also shown in CDRL. An explanation is that the DDRL scheme is not sensitive to the dynamic and  fast-changing wireless environment but CDRL needs more time steps to learn the representative features of the rapid-changing environment in high mobility scenarios, which results in a lower convergence. Note that the {DRL-based methods have} certain adaptability to environmental changes in user speed which can be interpreted as robustness on max Doppler frequency. Even though the correlation between adjacent channels is very small, the DRL-based frameworks still benefit from the exploration-exploitation strategy.
Similar results are also observed in \cite{zhang2021joint}.

In Fig. \ref{fig:rescheduling}, we assume that the users' rescheduling happens at the 50000th, 100000th, and 150000th-time slots. Instead of re-initializing the weights of all the DQNs in each agent, all of them continue the training process  based on  the designed information from {newly} scheduled users.
\textit{First},  a much higher start point and a comparable convergence time can be achieved in DDRL without witnessing a great performance collapse compared with the ZF-CI PCSI scheme. This can be interpreted by the fact that each agent tries to find the common features between the first scheduled and reschedule users which naturally makes a better decision based on these features and exhibits the ability for maintaining connectivity against user rescheduling in mobile networks. \textit{Second}, with more rescheduling happening, a higher information rate and a faster convergence can be observed in CDRL. An interpretation is that the common features learned from the previously scheduled users boost the training in rescheduling. After learning and 'storing' more and more feature information  into the weights of DQN, the central agent demonstrates
universality to the channel uncertainty of rescheduled users and the neural network weights extracted from the previously trained DQN is a good candidate for the weight initialization of the current trained DQN in both schemes. {However, if rescheduling happened frequently, the proposed DRL-based schemes can not converge to a set of good parameters if rescheduling happens before 25000-time slots but a jump-start happens when we implement a group of  trained DQN to new users. }
 
Fig. \ref{fig:users} investigates the average achievable information rate versus the number of time slots with $K = 4$ and $6$, respectively. As opposed to the decrease of average user rate, the total cell throughput improves which suggests that both DRL-based approaches can benefit from multi-user diversity provided by time-varying channels across  different users. In addition, although an 11.7$\%$ performance degradation can be observed in CDRL which is smaller than 19.8$\%$ in DDRL, CDRL achieves a much smaller performance gain over SAH in comparison to DDRL when $K = 6$. This suggests that DDRL is more robust in multi-user scenarios than CDRL. 

 {To sum up in a nutshell, on the one hand, CDRL and DDRL are not equally suitable for {massive} MIMO systems with channel aging, namely, CDRL is less computationally complex but demonstrates instability and incurs performance loss}. In contrast, DDRL offers a promising gain over CDRL by favoring an adaptive decision-making process and facilitating cooperation among all agents to mitigate interference but incurs a higher hardware complexity. {Also, DDRL is more robust on rescheduling than CDRL considering the convergence speed and stability. On the other hand, both schemes demonstrate robustness on fading characteristics of the environment and changes on  interference conditions.}
 
\subsection{Multi-antenna and Multi-stream}

\begin{figure}[t]
  \centering
    
    \begin{minipage}[b]{0.45\textwidth} 
    \centering 
    \includegraphics[scale = 0.5]{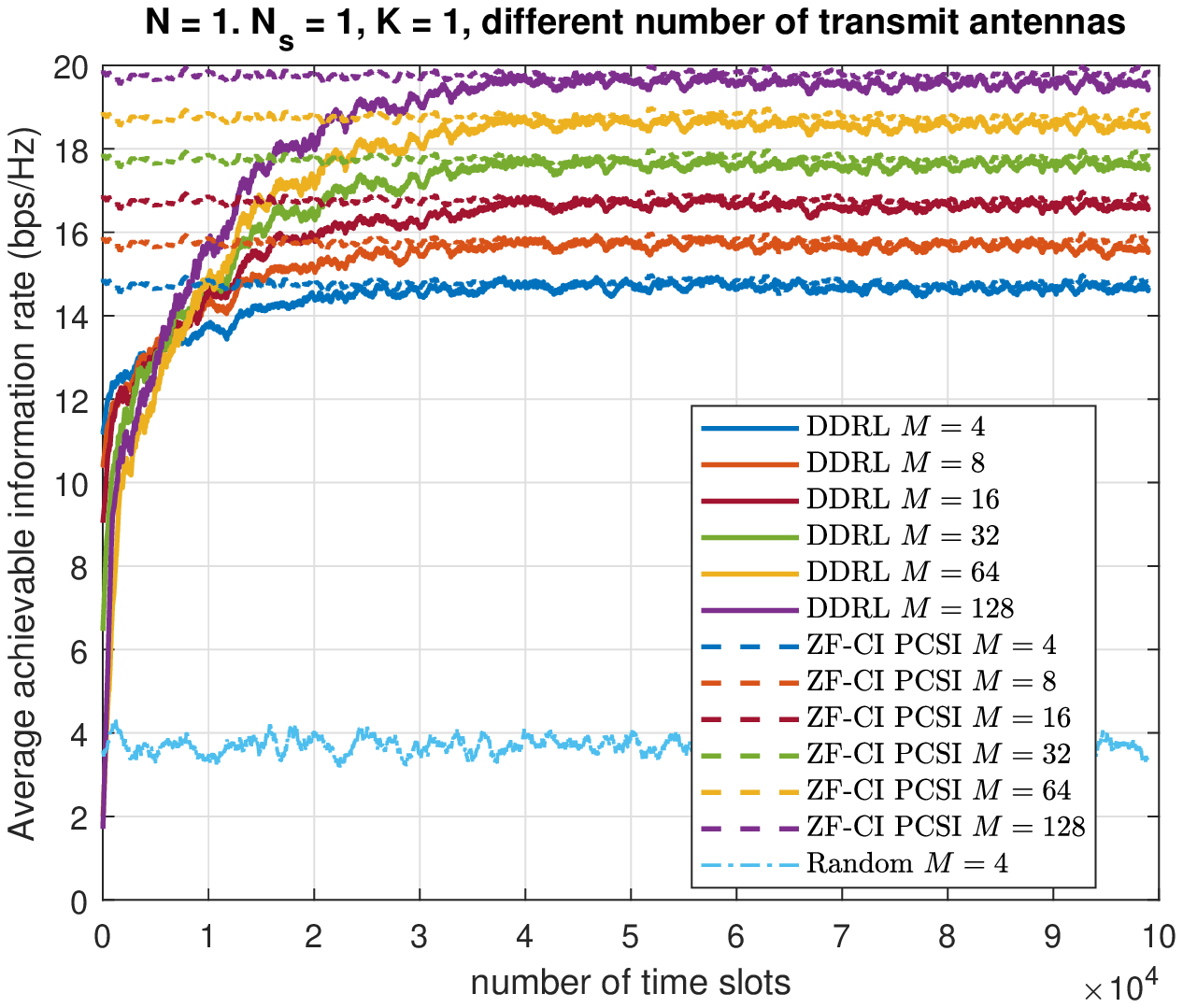}%
    
  \caption{Average information rate versus the number of time slots with different number of transmit antennas $M$ when $N = 1, N_s = 1, K = 1$.}
    \label{fig:single}
    \end{minipage}
    \begin{minipage}[b]{0.45\textwidth} 
    \centering 
    \hspace{-10mm}
    
  \includegraphics[scale = 0.5]{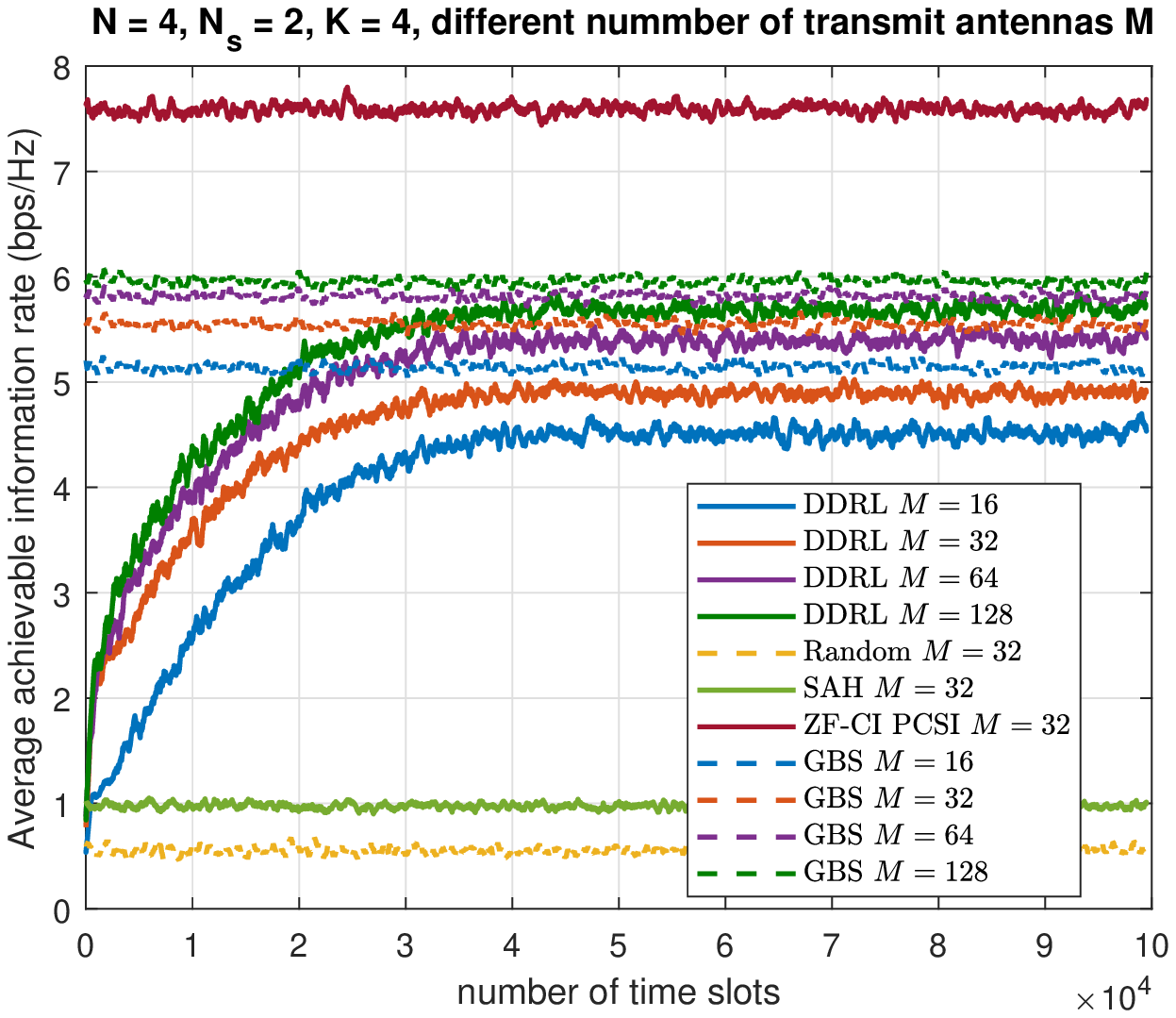}%
    
    \caption{Average information rate versus number of time slots with different number of transmit antennas $M$ when $N = 4, N_s = 2, K = 4$. }
   \label{fig:transmit}
   \end{minipage}
   
 \end{figure}
Fig. \ref{fig:single} illustrates the  DRL-based scheme with 6 different numbers of transmit antennas $M$ when $N = 1, N_s = 1, K = 1$. Without any penalty, i.e., inter-stream interference and multi-user interference, a near-optimal result can be observed by leveraging the proposed state, action, and reward design in an interference-free scenario with a stable increase of information rate, which thereby validates the effectiveness of the codebook design in Section \ref{section4}. {The convergence time is  proportional to $M$ which limits its scalability. Intuitively, the reason for this effect on the convergence time is that it takes a longer time to learn the representative feature of high-dimension CSIT in sequential states. } In contrast with Fig. \ref{fig:single},  a serious multi-user and inter-stream interference is managed in Fig. \ref{fig:transmit}  when $N = 4, N_s = 2, K = 4$. It can be observed that the transmit diversity and array gain cannot be fully achieved in the proposed DRL-based scheme if the rich interference is not properly suppressed due to the constraint of codebook precision and CSIT imperfections. Hence, the drawback of using DRL-based methods is that inter-stream interference can not be sufficiently alleviated if each agent fails to choose an action that causes small interference to all other agents during exploration and exploitation. 

\begin{figure}[t]
  \centering
    \begin{minipage}[b]{0.45\textwidth} 
    \centering 
    \includegraphics[scale = 0.5]{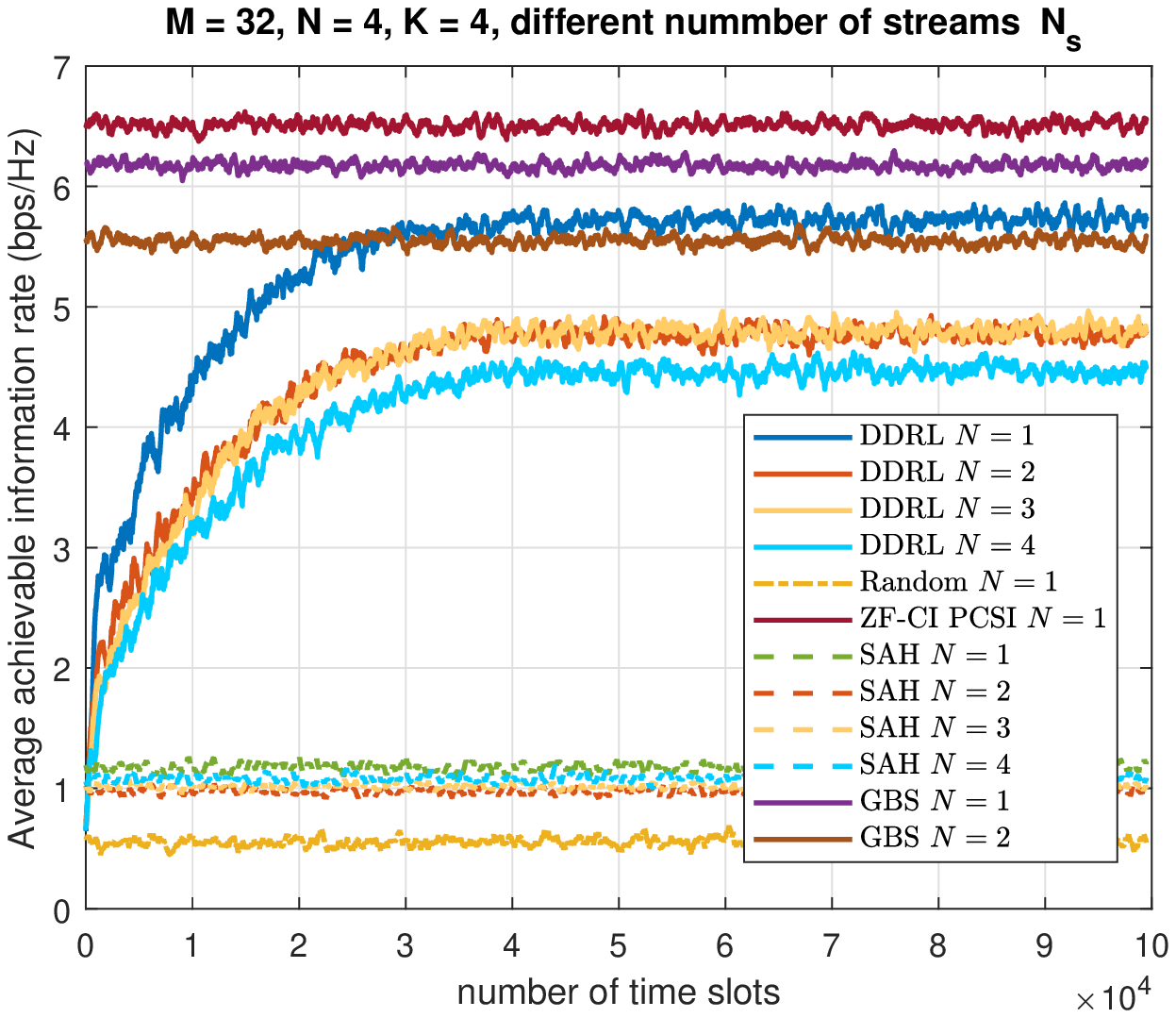}%
    
  \caption{Average information rate versus the number of time slots with different number of streams $N$.}
    \label{fig:nstreams}
    \end{minipage}
    \begin{minipage}[b]{0.45\textwidth} 
    \centering 
    \hspace{-10mm}
 
  \includegraphics[scale = 0.5]{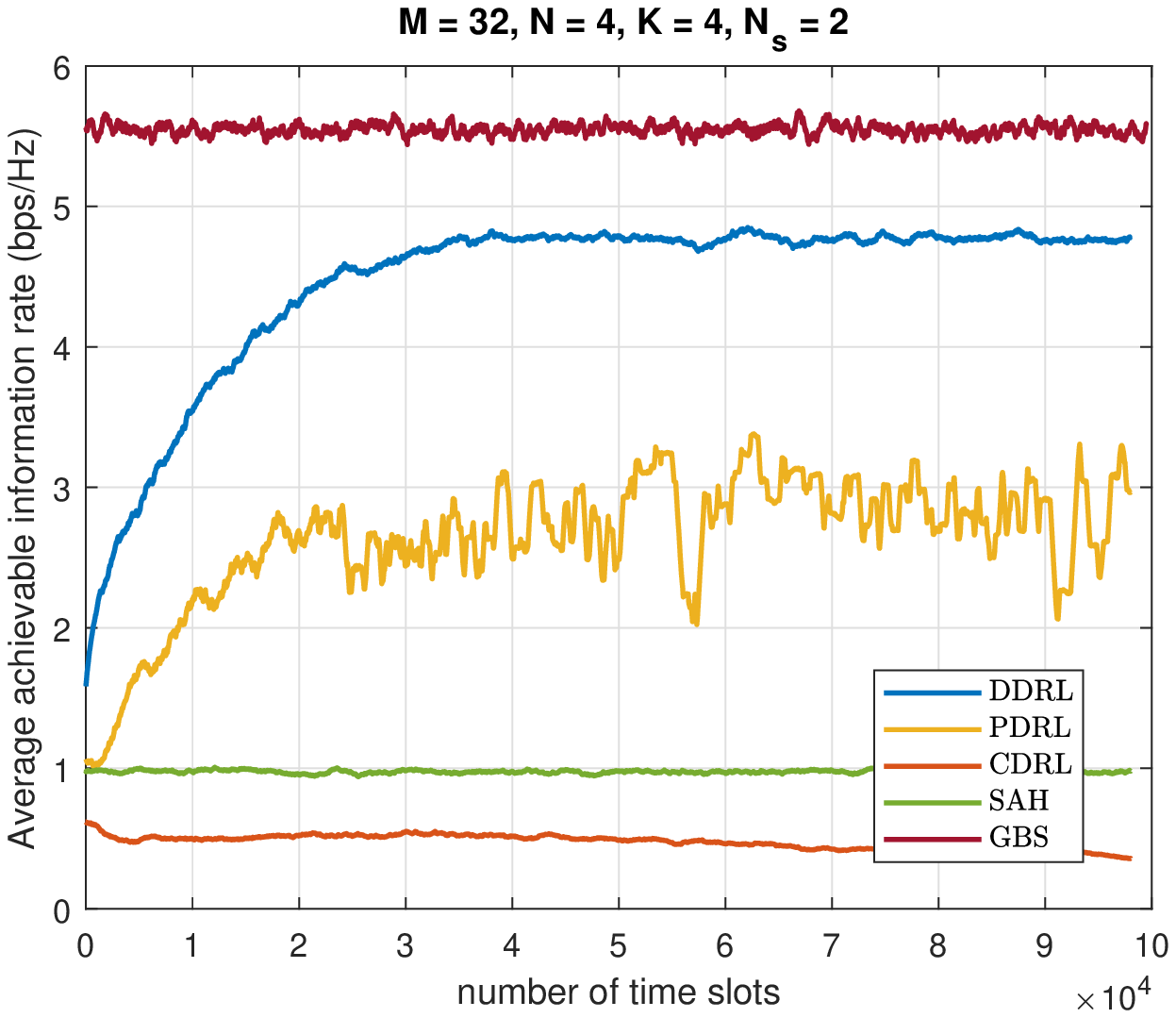}%
    \caption{Average information rate versus number of time slots with different DRL-based schemes.}
   \label{fig:partial}
   \end{minipage}
   
 \end{figure}
 
Fig. \ref{fig:nstreams} characterizes the average achievable information rate versus the number of time slots with different numbers of streams for each user.  First, DDRL has significantly higher performance compared to the conventional SAH scheme in different numbers of streams. Second, a 15.7$\%$  performance degradation can be observed from the DRL-based scheme between 1-stream and 2-stream scenarios which is smaller than that in SAH (around 28$\%$ between black and blue dotted lines). This reveals the privilege of the DDRL  in inter-stream interference management.

An overview of the average information rate versus number of time slots with three DRL-based algorithms is demonstrated in Fig. \ref{fig:partial}. Compared with DDRL and PDRL, a performance collapse is observed in CDRL due to the degrading effect of inter-stream interference.  By flexibly modeling each user as an agent, PDRL greatly mitigates the inter-stream interference by learning the local observations from the target user's propagation channel.  

\subsection{{Reward, Penalty Analysis and Statistical Test}}

\begin{figure}[t]
  \centering

    \begin{minipage}[b]{0.45\textwidth} 
    \centering 
    \includegraphics[scale = 0.5]{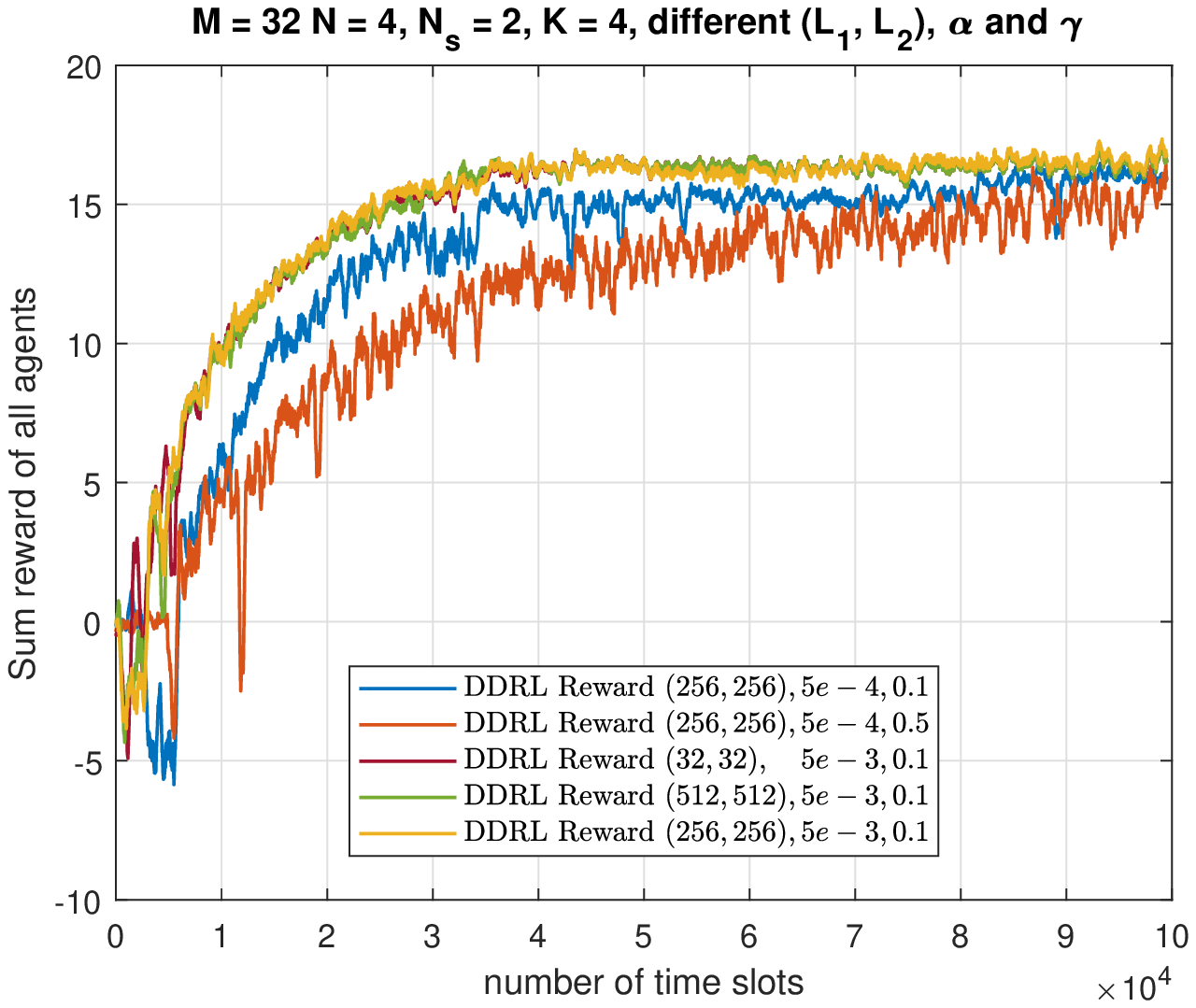}%
    
  \caption{Sum reward versus the number of time slots with different number of users $K$. }
    \label{fig:reward}
    \end{minipage}
    \begin{minipage}[b]{0.45\textwidth} 
    \centering 
    \hspace{-10mm}

  \includegraphics[scale = 0.5]{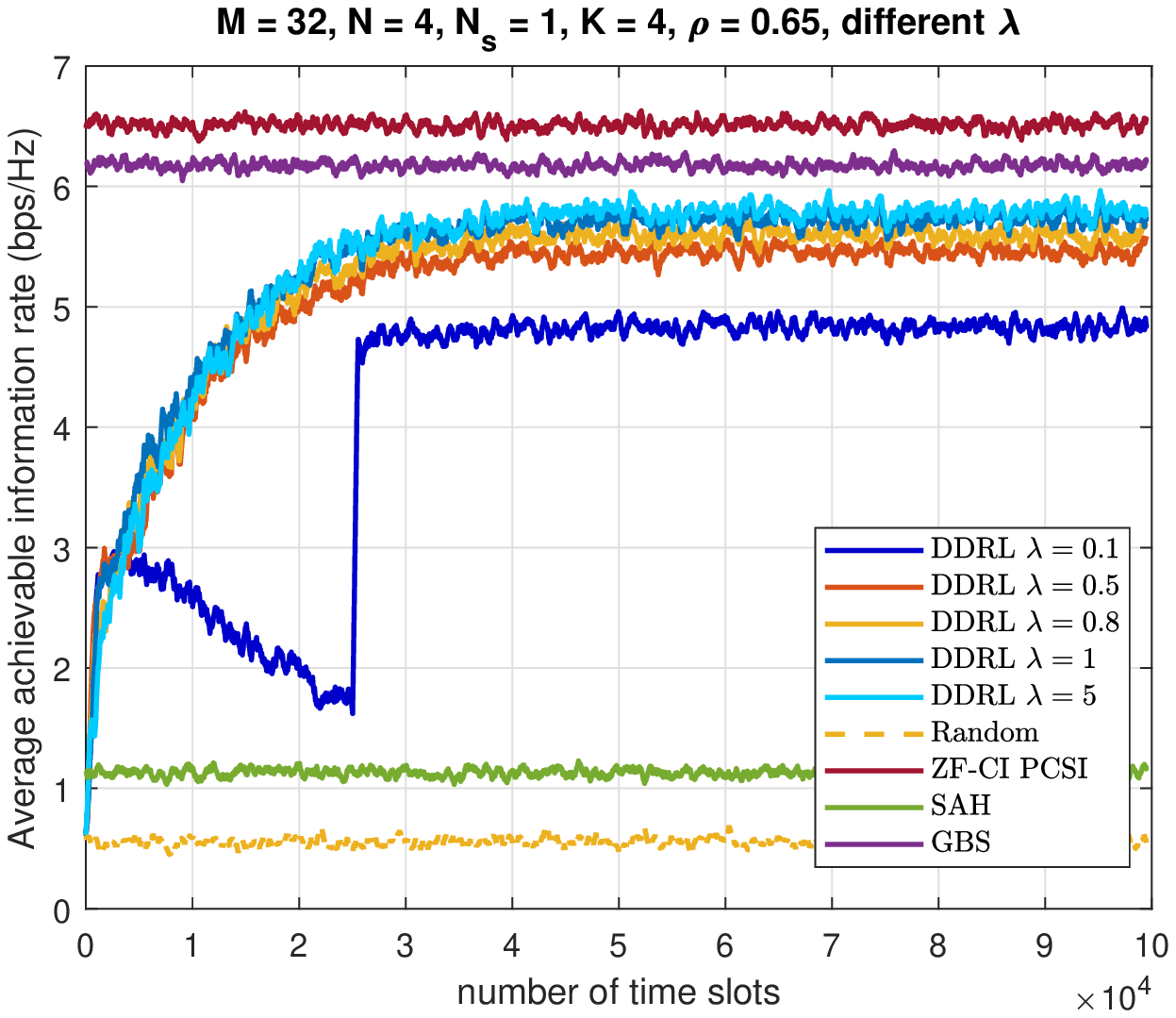}%
    
    \caption{Average information rate versus the number of time slots with different penalty $\lambda$. }
   \label{fig:penalty}
   \end{minipage}
   
 \end{figure}
 
To  reveal the significance of the neural network size ($L_1, L_2$), the learning rate $\alpha$ and the discount factor $\gamma$, Fig. \ref{fig:reward} shows the sum reward versus the number of time slots with different ($L_1, L_2$), $\alpha$ and $\gamma$. The first observation is that a faster convergence is observed with larger $\alpha$, this is intuitive since the  gradient descent is sped up with a larger value of loss function. The second observation is that, compared with (256, 256), a reward degradation appears with (32, 32), which suggests that increasing the DNN size demonstrates a stronger representation capability of input features and boosts the performance of the DRL-based scheme.  Due to the negligible performance improvement in (512, 512), (256,256) is chosen as a baseline to maintain a balance between user connectivity and computational burden.

Fig. \ref{fig:penalty} offers an insight into the impact of different penalty values $\lambda$. This penalty term   intrinsically represents an adjustment of reward function for each agent. Different from  \cite{nasir2019multi, ge2020deep}, it is demonstrated in Fig. \ref{fig:penalty} that the system capacity is gradually increased with the penalty value from 0.1 to 5. An interpretation is that each agent causes high interference to other agents while still trying to maximize its  information rate. Due to the uncertainty of the dynamic environment, the lack of perfect CSI introduces unpredictable  interference for all the agents and an increase of penalty value can make a remedy for this by choosing an action that minimizes the 
interference to other agents instead of maximizing the received power of itself. This result also indicates that the decision-making process of all the agents is robust in unexpected high-interference scenarios.
\begin{figure}[t]
  \centering

    \begin{minipage}[b]{0.45\textwidth} 
    \centering 
    \includegraphics[scale = 0.5]{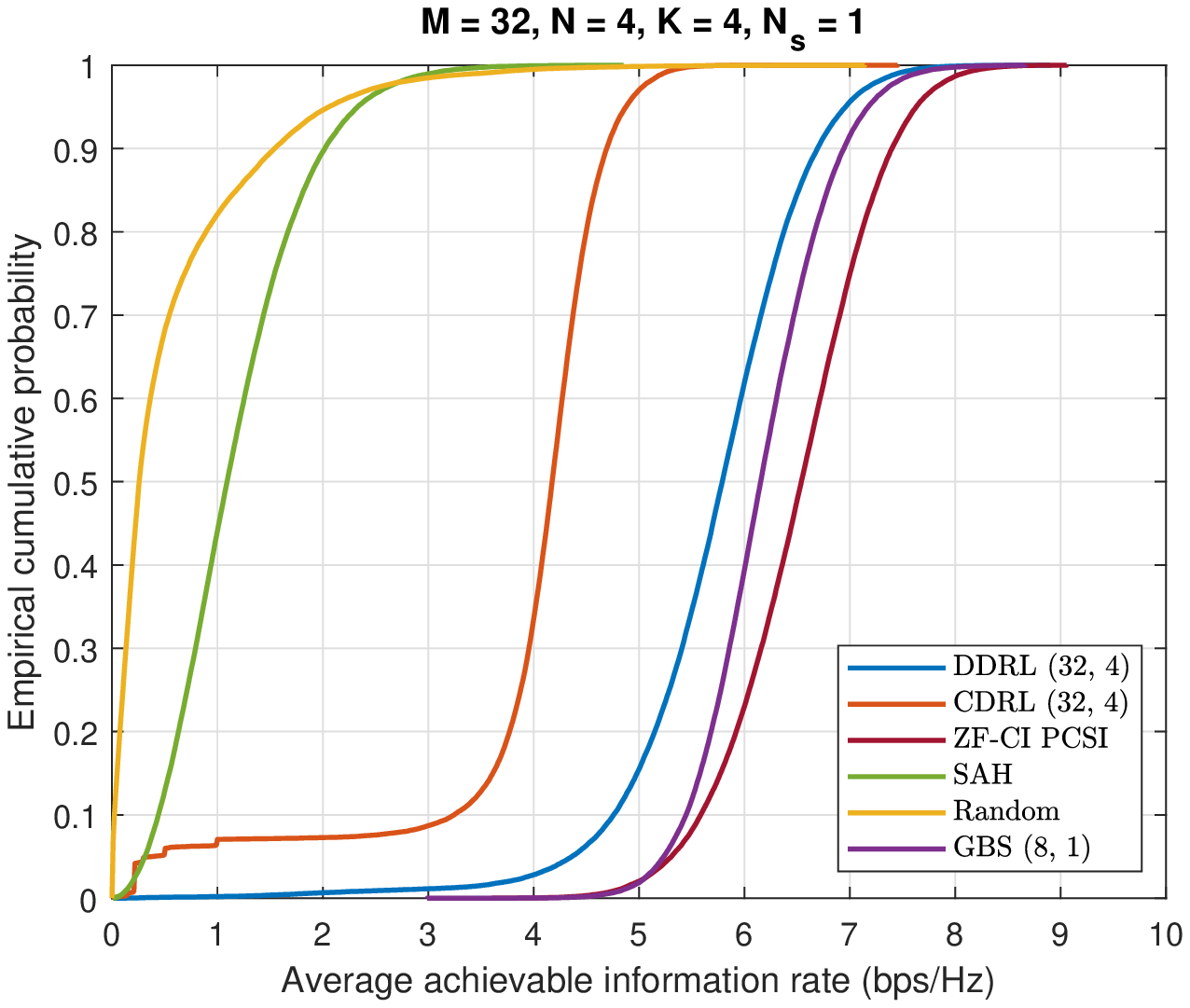}%
    
  \caption{{Cumulative distribution function (CDF) of the average information rate over different DRL-based methods and benchmarks.} }
    \label{fig:CDF}
    \end{minipage}
    \begin{minipage}[b]{0.45\textwidth} 
    \centering 
    \hspace{-10mm}

  \includegraphics[scale = 0.5]{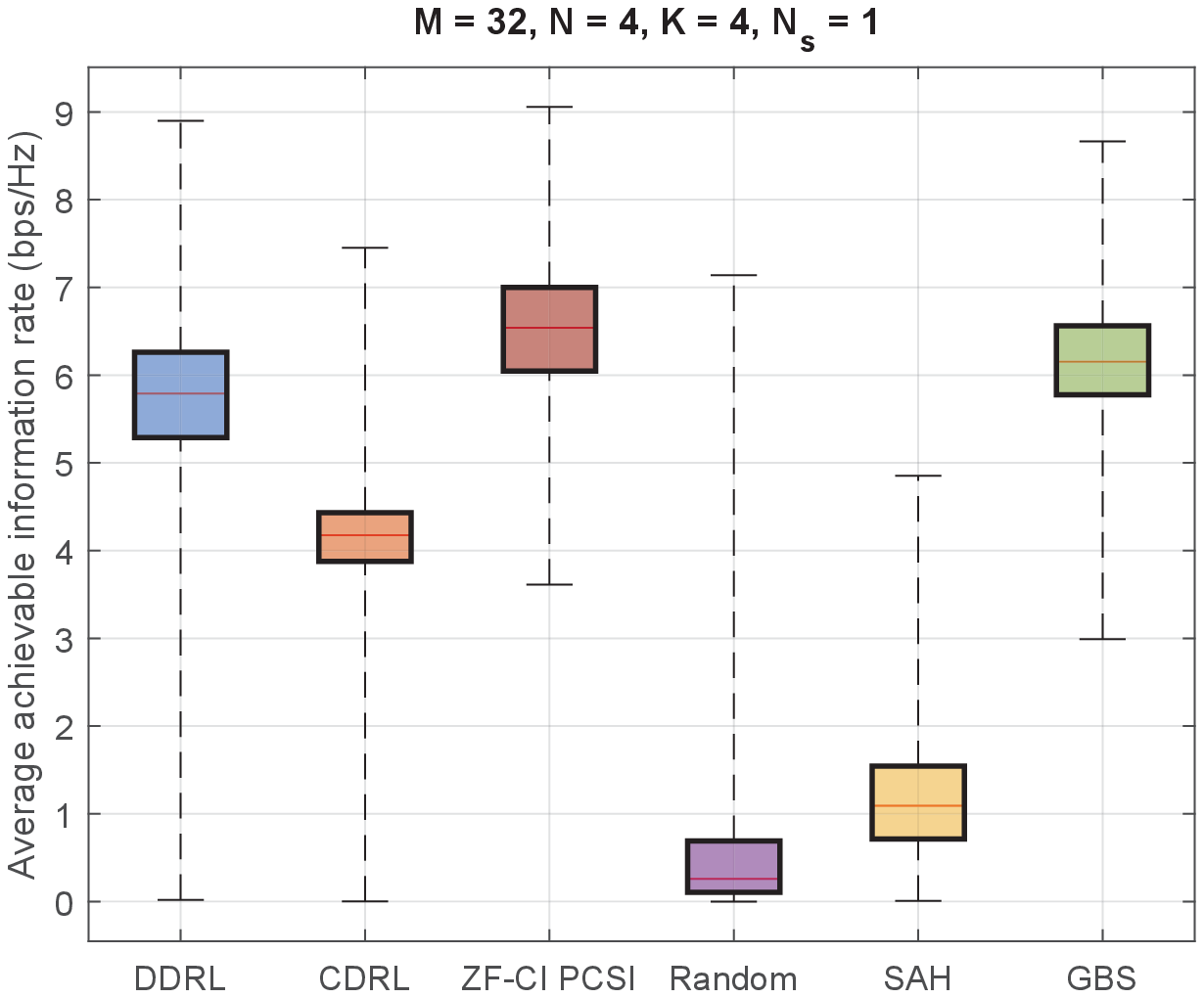}%
    
    \caption{{Boxplot of the average information rate over different DRL-based methods and benchmarks. The bottom and top of each box are the 25th and 75th percentiles of the information rate values, respectively. The whisker length is set to infinity to ensure there are no outliers. }}
   \label{fig:boxplot}
   \end{minipage}
   
 \end{figure}
 
{The cumulative distribution function (CDF) over different DRL-based methods and benchmarks have been plotted in Fig. \ref{fig:CDF}. It can be seen that the CDF curves confirm  the discussion of the superiority of DDRL over other schemes. Also, the performance of DDRL is significantly limited by the codebook resolution which is confirmed in Fig. \ref{fig:boxplot} where we show the boxplot of different methods.}

\section{Conclusion {and Future Research}}\label{section6}
{In this paper, we studied the beamforming optimization for massive MIMO downlink transmission with channel aging.  An optimization framework in light of DRL was studied and three DRL-based algorithms were derived based on stream-level, user-level, and system-level agent modeling. Specifically, the transmit precoder at BS and receive combiner at user terminals were jointly optimized to maximize the average information rate. Furthermore, we analyzed the performance loss of DRL-based approaches as compared to the ideal case with continuous beamforming with different numbers of codebook sizes, users, antennas, streams, and user speeds. Interestingly, it was shown that even using a very low-resolution codebook in DDRL is still able to achieve 95$\%$ and $90\%$ as in the case with GBS and ZF-CI, respectively. Simulation results showed that significant robustness on user mobility can be achieved by using some received power values of imperfect CSIT at the expense of more uplink overhead. Also, the convergence speed and scalability of the proposed algorithms are discussed. The convergence speed is linearly increased with the number of transmit antennas and performance degradation in the multiuser case is non-negligible due to the severe co-channel interference. In addition, CDRL consumes less computation complexity but demonstrates instability and incurs performance loss. In contrast, DDRL offers a promising gain over CDRL by favoring an adaptive decision-making process and facilitating cooperation among all agents to mitigate interference but incurs a higher hardware complexity and non-stationarity. Finally, the  reward, penalty analysis, and statistical test confirm the fact that the performance of the proposed algorithms is greatly limited by the resolution of codebooks. 
}
{Several important issues that are not addressed in our paper yet, some of which are listed as follows to motivate future research. 
\begin{itemize}
    \item \textit{Multi-cell}: This paper considered single-cell multiuser conditions. However, when multi-cell is considered. The transmit power of the BS  needs to be optimized, due to which the corresponding optimization problem is more challenging to solve, and thus is worthy of further investigation.
    \item \textit{Extremely Large-scale MIMO}: To overcome the capacity constraints of conventional MIMO, extremely large-scale MIMO (XL-MIMO) are  being proposed which can provide a much stronger beamforming gain to compensate for the severe path loss. As such, it is worth comparing the  proposed massive MIMO with the XL-MIMO in future investigations.
    \item \textit{Beamforming Codebook Design}:  As is shown in this paper, the performance is greatly limited by the resolution of the designed codebook. A better codebook enables the system to handle larger and more complex channel conditions without compromising on performance.
\end{itemize}
}


\appendices
\section{PDRL Algorithm}
\begin{algorithm}[h]
    \caption{PDRL  Algorithm}\label{alg:PDRL_MU_MIMO}
    \begin{algorithmic}[1]
        \State{$\textbf{Initialize}$: Establish $K$ pairs of trained/target  DQNs with random weights $\bm{\theta}_{k}$ and $\bm{\bar{\theta}}_{k}, \forall k \in \{1, 2, 3, \dots, K\}$ as user-specified agents, update the weights of $\bm{\bar{\theta}}_{k}$ with random $\bm{\theta}_{k}$. Build experience pool  $\mathcal{O}_{k}, \forall k$.
        Establish a  trained  DQN with weight $\bm{\theta}_{k,n}$, $\forall k \in \left\lbrace 1, 2, \ldots, K \right\rbrace, \forall n \in \left\lbrace 1, 2, \ldots, N \right\rbrace$ for each distributed agent.}
        
        \State{In the first $E_s$ time slots,  agent $(k, n)$ randomly selects an action from action space $\mathcal{A}$, and stores the  experience $\langle s_{k,n}, a_{k,n}, r_{k,n}, s_{k,n}^{\prime}\rangle, \forall k, n$ in  the experience pool of corresponding  user-specified agent  $\mathcal{O}_{k}$.}
        
        \State{$\textbf{for}$  each time slot $t$ $\textbf{do}$}
             \State{\quad$\textbf{for}$ each agent $(k, n)$ $\textbf{do}$}
            \State{\quad\quad Obtain state $s_{k, n}$ from the observation of agent $(k, n)$.}   
            \State{\quad\quad Generate a random number $\omega$.}
            \State{\quad\quad\textbf{If} $\omega < \epsilon$ $\textbf{then}$:}
            \State{\quad\quad\quad Randomly select an action in action space $\mathcal{A}$.}
            \State{$\quad\quad\textbf{Else}$}
            \State{\quad\quad\quad Choose the action $a_{k,n}$ according to the Q-function $q(s_{k,n}. a; \bm{\theta}_{k,n}), \forall k, n$ }
            \State{\quad\quad\textbf{End if }.}
        \State{\quad\quad Agent $(k, n)$ executes the  $a_{k,n}$, immediately receives the reward $r_{k, n}$ and steps into next state $s_{k, n}^{\prime}, \forall k, n$.}
        \State{\quad\quad Agent $(k, n)$ puts experience $\langle s_{k,n}, a_{k,n}, r_{k,n}, s_{k,n}^{\prime}\rangle$ into central experience pool $\mathcal{O}_{{k}}$.}
        \State{\quad\textbf{\textbf{end for}}}
        \State{\quad User-specified agent $k$  randomly samples a minibatch with size $E_b$. Then, the weights of its trained DQN $\bm{\theta}_{k}$ are updated using back propagation approach. The weights of its target DQN $\bm{\bar{\theta}}_{k}$ is updated every $T_s$ steps. Then, the user-specified agent broadcasts the weights $\bm{\theta}_{k}$ to  the corresponding distributed agents, i.e., $\bm{\theta}_{k,n} = \bm{\theta}_{k}, \forall  n.$}
        
        \State{\textbf{\textbf{end for}}}
    \end{algorithmic}
\end{algorithm}

\section*{ACKNOWLEDGMENT}
The author would like to thank Hongyu Li, Yumeng Zhang, and Dr. Onur Dizdar for stimulating discussions.


\bibliographystyle{IEEEtran} 
\bibliography{main.bib}
\end{document}